\documentclass[11pt]{article}

\usepackage[margin=1in]{geometry}
\usepackage{amsmath,amssymb,amsthm}
\usepackage{physics}
\usepackage{tikz}
\usepackage{quantikz}
\usetikzlibrary{matrix,positioning}
\usepackage{hyperref}
\usepackage{booktabs}
\usepackage{graphicx}
\usepackage[markup=bfit,deletedmarkup=sout,authormarkup=none]{changes}
\definechangesauthor[color=red]{Paulet}
\definechangesauthor[color=brown]{FLP}
\definechangesauthor[color=magenta]{HICC}
\definechangesauthor[color=orange]{FPG}

\usepackage[normalem]{ulem}
\usepackage{xcolor}
\usepackage{algorithm}%
\usepackage{algpseudocode}%
\usepackage{listings}%

\newtheorem{theorem}{Theorem}
\newtheorem{example}{Example}

\newtheorem{lemma}{Lemma}
\newtheorem{definition}{Definition}

\title{A Grover-Amplified Oracle for Maximum Clique on Graph, Minimum Hitting set and Dualizations on Hypergraphs Built on Quantum Phase Estimation}

\author{
  Francisco Pi\~nero \and Fernando L.\ Pelayo \and Hernan I.\ de la Cruz \and Jose J.\ Paulet \and Mauro Mezzini \and Fernando Cuartero 
}
\date{}

\begin{document}

\maketitle

\begin{abstract}
In this paper, we use a quantum circuit called the Graph Encoding Operator which, applied to the complement $\bar G$ of any graph $G$, yields an oracle for the maximum clique problem suitable for amplification via Grover's algorithm.
Analogously, we devise a Hypergraph Encoding Operator which, when applied to a hypergraph $\mathcal H$, yields an oracle for the minimum hitting set suitable for amplification via Grover's algorithm. Interestingly, the latter oracle provides, when measured, an approximation of the number of hitting sets of the hypergraph, thus solving the dualization problem for positive Boolean functions represented in a positive irredundant disjunctive normal form.
We give the full construction of both operators, prove their correctness, analyze their complexity, and provide an implementation on a quantum simulator. 
We discuss the resulting complexity, contrast them with existing quantum computer algorithms based on Grover search and we will show that our algorithms outperform, to the best of our knowledge,  all of the previous algorithms on space computational complexity by requiring only  $O(N)$ number of qubits with respect to the number of vertices $N$, for representing both graphs and hypergraphs in the encoding operator and, consequently, results in an oracle for both graphs and hypergraphs that uses only $O(N)$ qubits even when the hypergraph contains an exponential (on the number of vertices) number of hyperedges.
\end{abstract}
\medskip

\section{Introduction}\label{sec:intro}

Finding a maximum clique in a graph $G=(V,E)$ --- the largest subset of pairwise-adjacent vertices --- is one of the classical NP-hard problems~\cite{Karp1972}, and it remains hard to approximate within polynomial complexity in the worst case. Its complementary formulation, the maximum independent set problem, is equally hard since they both are related by the standard duality: $S$ is a clique in $G$ if and only if $S$ is an independent set in the complement graph $\bar G$ of $G$. Furthermore the  minimum vertex cover problem and its generalization to hypergraphs, that is the minimum hitting set problem, are also very well known classical NP-hard problems~\cite{Karp1972} strictly related to maximum clique and maximum independent set problems.  

The Maximum Clique Problem (MCP) has applications in various scientific and engineering disciplines and it is fundamental in fields such as medicine, computational biology, coding theory, economics, financial market analysis, telecommunications, wireless networks, cryptography, cybersecurity, and physics \cite{okamoto2020finding,rozman2024enhanced,boginski2005statistical,douik2020tutorial,lu2023atomic}. Furthermore, its structural connections with other combinatorial optimization problems, such as graph coloring and clustering, ensure that advances in solving the MCP translate directly into broader algorithmic breakthroughs. Given its fundamental importance, developing more scalable algorithms for the MCP remains a primary challenge, motivating the exploration of new paradigms in quantum computing.

Quantum approaches to clique-related problems have been proposed  using both the gate-based quantum computing and quantum-annealing / QUBO paradigms. 
This paper will use the Graph Encoding Operator (GEO) of our companion paper~\cite{PelayoQPE2024} --- originally designed as a graph isomorphism invariant --- to develop a Grover oracle. Such an operator computes  the number of edges of the subgraph induced by an arbitrary vertex subset $S$; a clique is exactly a subset inducing the empty graph in $\bar G$, so the very same circuit run on $\bar G$ instead of on $G$ and left uncollapsed, already tells us --- coherently, in superposition --- which subsets are cliques.

By employing a modification  of the GEO we were able to devise a Hypergraph Encoding Operator (HEO) capable of counting the number of hyperedges covered, that is, that have a non empty intersection to an arbitrary subset of vertices thus, giving the information on which subset of vertices represent a hitting set, since a hitting set covers all the hyperedges of the graph. As in the graph counterpart this operator could be employed as Grover oracle.    
Interestingly, both the oracles based on GEO and  HEO require only $N+\lceil \log_2  M \rceil+\lceil \log_2  N \rceil+3$ qubits where $N$ and $M$ are, respectively the number of vertices  and the number of (hyper)edges of the (hyper)graph. Therefore, in the case of a graph where the number of edges belongs to $O(N^2)$, the number of qubits required belongs to $O(N)$. To the best  of our knowledge~\cite{Marino2026} this compares favorably to all existing  Grover based quantum algorithms for the clique problem which the best computational space complexity belongs to $O(M)$ qubits. Thus our proposal decreases dramatically the space requirements for solving the maximum clique problem using Grover iteration with respect to the existing algorithms which belongs to $\Omega(N^2)$. 
Furthermore, to the best of our knowledge, the HEO is the first quantum based oracle suitable for Grover iteration in order to solve the minimum hitting set problem in hypergraphs and, even when the number $M$ of hyperedges is exponential in the number of vertices, that is, when $M \in O(2^N)$, our oracle requires $O(N+\log_2 2^N)= O(N)$ qubits. Moreover, much interestingly, the single HEO can, when measured, approximate the number of hitting sets of the hypergraph; and in turn, finding the  number of hitting set of the hypergraph directly solves  the problem of dualization of positive boolean function represented by a positive irredundant disjunctive normal form \cite{Mezzini2024}.

The paper is organized in the following way. In Section \ref{sec:prelim}, we give some preliminaries and recall the description of the GEO from \cite{PelayoQPE2024}. In Section \ref{sec:counting-to-marking}, we describe the maximum clique/independent set oracle, prove its correctness, describe the space and gate complexities and check the numerical results from our implementation of a full Grover algorithm. In Section \ref{sec:MHSP_dualization}, we describe the HEO, prove its correctness, analyze the space and gates complexities and show how to obtain an approximation on the number of hitting set of a hypergraph for solving the dualization problem of positive Boolean function. In addition to this, the Quiskit code is provided to  build the GEO and HEO oracles. In Section \ref{sec:discussion}, we discuss about related works and compare their space and time computational complexities with our approach. Finally, in the Conclusion section we give some concluding remarks and point at possible future works.

\section{Preliminaries}\label{sec:prelim}

A \textit{graph} is a pair $G=(V(G), E(G))$ where $V(G)$ is a set of vertices and $E(G)$ is a set of subsets of $V(G)$ having exactly two 
vertices. We often denote an edge $\{u,v\}$ as $uv$. We will denote $N=|V(G)|$ and, without loss of generality, we identify the set $V(G)=\{0,1,\dots, N-1\}$.  
A graph $G$ is said to be \textit{complete} if for every pair of distinct vertices $u, v \in V(G)$ we have that $uv \in E(G)$. 
A graph $\bar G$ is the \textit{complement} of a graph $G$ if $V(\bar G)=V(G)$ and $E(\bar G)= \{ uv | u,v \in V(G) \wedge uv \notin E(G)\ \wedge \ u \neq v \}$.
If $G$ is a graph and $S\subseteq V(G)$, the subgraph of $G$ \textit{induced} by $S$ is the graph denoted as $G_S$ where $V(G_S)=S$ and $E(G_S)= \{ uv \in E(G)| u \in S \wedge v  \in S\}$. A \textit{clique} of $G$ is a subset  $S \subseteq V(G)$ such that $G_S$ is a complete graph. An \textit{independent} set $S$ of $G$ is a subset of $V(G)$ such that $E(G_S)= \emptyset$.
A \textit{vertex cover} of $G$ is a subset $S \subseteq V(G)$ such that $S \cap \{u,v\}\neq \emptyset$ for all $uv \in E(G)$. The size of a maximum  clique in a graph $G$ is called the \textit{clique number} of $G$ and is denoted by $\omega(G)$.
The \textit{Maximum Clique Problem} (MCP) asks to find a clique of maximum size in a graph. The \textit{Maximum Independent Set Problem} (MISP) asks to find an independent set of maximum size in a graph. The \textit{Minimum Vertex Cover Problem} (MVCP) asks to find a vertex cover of minimum size in a graph.  The decision variant of the MCP (resp. MISP) asks, given an integer $k$, whether there exists a clique (resp. an independent set) $S$ in a graph such that $|S|\geq k$. For the MVCP, the decision counterpart asks given an integer $k$, whether there exists a vertex cover $S$ in the graph such that $|S|\leq k$.

A \textit{hypergraph} $\mathcal{H}=(V(\mathcal{H}), E(\mathcal{H}))$  is a generalization of a graph, where $V(\mathcal{H})$ is a set of vertices and $E(\mathcal{H})$ is a set of non-empty subsets of $V(\mathcal{H})$ called \textit{hyperedges}.
The hypergraphs we will deal with have the property that $V(\mathcal{H})=\bigcup_{\mathcal {E} \in E(\mathcal{H})} \mathcal {E}$ so giving the set of hyperedges will implicitly define the vertex set, and consequently we identify $\mathcal{H}$ with its set of hyperedges. 
Given a $v \in V(\mathcal{H})$ then $N_{\mathcal{H}}(v)$ denotes the set of hyperedges containing $v$, that is $N_{\mathcal{H}}(v)= \{ \mathcal {E} \in E(\mathcal{H})| v \in \mathcal {E}\}$.
Given $S \subseteq V(\mathcal{H})$, we denote by $E_S$ the set $E_S=\{\mathcal {E} \in E(\mathcal{H})| \mathcal {E} \cap S \neq \emptyset\}$. We say that $S$ \textit{covers} $E_S$. If $E_S=E(\mathcal{H})$ then $S$ is a \textit{hitting} set of $\mathcal{H}$. Note that if $v$ is a vertex of $\mathcal{H}$ then $E_{\{v\}} = N_{\mathcal{H}}(v)$.
The problem of finding a minimum hitting set of a hypergraph $\mathcal{H}$ (MHSP) is NP-hard by restriction to the MVCP in graphs~\cite{Karp1972}. Its decision counterpart asks, given a natural number $k$, whether there exists a hitting set $S$ in $\mathcal H$ such that $|S|\leq k$.

We will also make use of the following notation. We will consider $x$, satisfying  $0 \leq x < 2^N$, sometimes as a number and sometimes as its $N$-bit binary representation; with $S(x)$ we will denote the subset of $\{0,1,\dots, N-1\}$ whose characteristic vector is $x$. For example, if $N=5$ and $x =13 = (0,1,1,0,1)$ then $S(x)=\{1,2,4\}$. 
\\

We recall the definitions and the central theorem from~\cite{PelayoQPE2024} that this paper builds on. We will build our oracles by using the (multi-controlled) \textit{phase gate} $P(\theta)$:

\begin{equation} \label{eq:phase-gate}
    P( \theta) =\begin{bmatrix}
1 & 0 \\
0 & e^{i \theta}
\end{bmatrix}
\end{equation}

\begin{lemma} \label{le:phase-gate}
    We have that $P(\theta)\ket{0}=\ket{0}$ and $P(\theta)\ket{1}=e^{i\theta}\ket{1}$.
\end{lemma}

\begin{definition}[cf.\ Def.~4.3 of~\cite{PelayoQPE2024}]\label{def:UGbar}
Let $G$  be a graph and $\theta \in (0,2\pi]$. For $uv \in E(G)$, let $A^\theta_{uv}$ be the operator applying the phase gate $P(\theta)$ on qubit $u$, controlled by qubit $v$, acting as the identity elsewhere. We define
\[
U_{G}^\theta = \prod_{uv \in E(G)} A^\theta_{uv} 
\]
as the \textit{Graph Encoding Operator} (GEO) of $G$.
\end{definition}

\begin{lemma}[cf. Lemma 4.2 of~\cite{PelayoQPE2024}]\label{le:phase-operator}
    If $\ket{x}$ is a vector of the computational basis whose binary representation is $\ket{x}=\ket{x_0x_1...x_{N-1}}$ then 
\begin{equation*}
A^\theta_{uv}\ket{x} = \begin{cases}
e^{i\theta}\ket{x} & \text{ if } x_u = x_v = 1\\
\ket{x} & \text{otherwise }
\end{cases}
\end{equation*}
and thus any computational basis state $\ket{x}$ is an eigenvector of $A_{uv}^{\theta}$.
\end{lemma}
\noindent
From Lemma \ref{le:phase-operator} we  immediately derive the  following Lemma:
\begin{lemma}[cf. Proposition 4.4 of~\cite{PelayoQPE2024}]
Any computational basis state $\ket{x}$ is an eigenvector of $U_G^{\theta}$ with the corresponding eigenvalue $e^{i\theta|E(G_{S(x)})|}$.
\end{lemma}
\noindent
From these observations we recall the following theorem.
\begin{theorem}[Theorem~4.5 of~\cite{PelayoQPE2024}]\label{thm:qpe-count}
Let $QPE_{G}$ be the quantum phase estimation circuit built from $U_{G}^\theta$ having one register of $N$ qubits for the GEO of $G$, a phase register of $m=\lceil \log_2|E(G)|\rceil+1$ qubits for the phase estimation and $\theta = 2\pi/2^{m}$. Then
\[
QPE_{G} \left( \ket{0}^{\otimes m}\otimes \ket{x} \right)\;=\;  \ket{\ |E(G_{S(x)})| \ }\otimes \ket {x}
\]
\end{theorem}
\noindent
In other words, Theorem \ref{thm:qpe-count} states that, for every $x$ such that $0 \leq x<  2^{N}$, i.e., for every $S(x) \subseteq V(G)$, the phase estimation register ends up in the basis state encoding $|E(G_{S(x)})|$, the number of edges of the subgraph of $G$ induced by $S(x)$. Note that if $|E(G)|$ is not a power of two we can save one qubit in the phase estimation register, that is, we can use $m=\lceil \log_2|E(G)|\rceil$ qubits. \\

Now, $S$ is an independent set of a graph $G$ if and only if $|E(G_S)|=0$ and therefore, we can leverage the phase register to build a Grover oracle: all we need is a circuit that tells us whether $|S|\geq k$ for some fixed integer $k$. We will see that this can be done using $\lceil \log_2 N \rceil + 1$ additional qubits. Clearly if $S$ is an independent set of cardinality greater than or equal $k$ then $V(G) \setminus S$ is a vertex cover of cardinality less than or equal to $N-k$. 
Furthermore, if we search for an independent set of cardinality greater than or equal  $k$ in the complement graph $\bar G$ of $G$ we effectively have got an oracle for the decision version of the MCP of $G$.

\subsection{Worked example: graph encoding and phase estimation}\label{sec:worked-example}

\begin{figure}[htbp]
\centering
\includegraphics[width=0.3 \textwidth ]{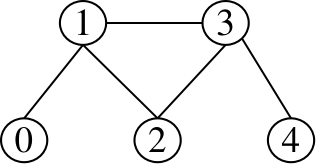}
\caption{A graph $G$ isomorphic to its complement.}
\label{fig:graph-example}
\end{figure}

Throughout this section we illustrate every construction on a single running example. As we mentioned before, the method always involves two related graphs: the target graph $G$, whose maximum clique we seek, and its complement $\bar G$, the one actually encoded into the circuit (a subset $S\subseteq V(G)$ is a clique of $G$ if and only if it induces no edges of $\bar G$). For this example, we use the graph with the following edge set (see Figure~\ref{fig:graph-example}):

\[
E(G)=\{\{0,1\}, \{1,2\}, \{1,3\}, \{2,3\}, \{3,4\}\}
\]
In this instance, its complement $\bar G$ happens to be isomorphic to $G$ itself, so a single drawing suffices for both. Consequently, in the example that follows, we can reverse their roles—treating $\bar G$ as the original graph and $G$ as its complement—while continuing to base our analysis on $G$.

\subsubsection{Graph encoding circuit}\label{sec:encoding-circuit}

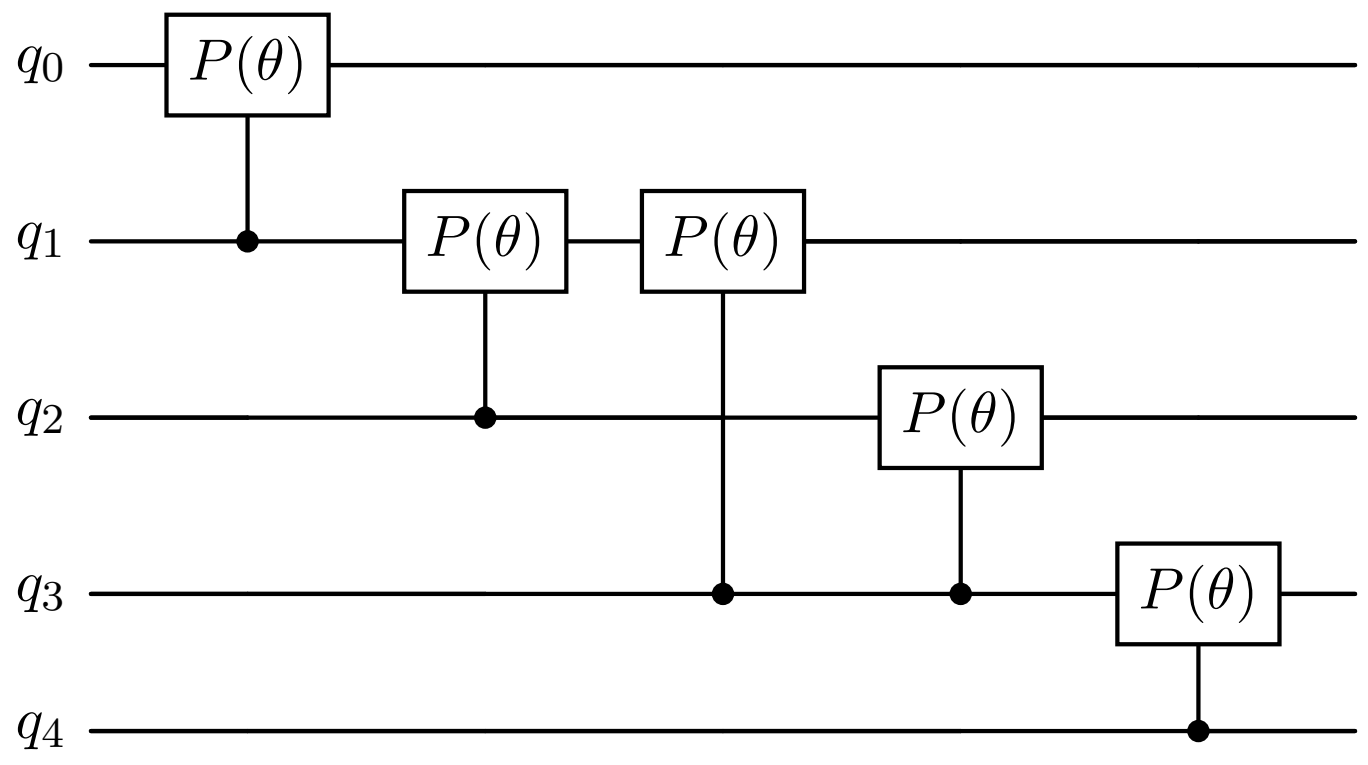
\begin{figure}[htbp]
\centering
\begin{quantikz}
\lstick{$q_0$} &  \gate{P(\theta)}   &  \qw         & \qw       & \qw       & \qw       & \qw\\
\lstick{$q_1$} &  \ctrl{-1}  &  \gate{P(\theta)}    & \gate{P(\theta)}  & \qw       & \qw       & \qw\\
\lstick{$q_2$} &  \qw        &  \ctrl{-1}   & \qw       & \gate{P(\theta)}  & \qw       & \qw\\
\lstick{$q_3$} &  \qw        &  \qw         & \ctrl{-2} & \ctrl{-1} & \gate{P(\theta)}  & \qw\\
\lstick{$q_4$} &  \qw        &  \qw         & \qw       & \qw       & \ctrl{-1} & \qw
\end{quantikz}
\caption{The GEO $U_{G}^{\theta}$ for the worked example.}
\label{fig:encoding}
\end{figure}

Figure~\ref{fig:encoding} shows the GEO of the graph of Figure~\ref{fig:graph-example}, following the convention of ~\cite{PelayoQPE2024}: each column corresponds to an edge $uv$, with $u < v$, and contains a phase gate $P(\theta)$ applied to qubit $u$ and controlled by qubit $v$.


\subsubsection{Phase estimation}\label{sec:phase-estimation-circuit}

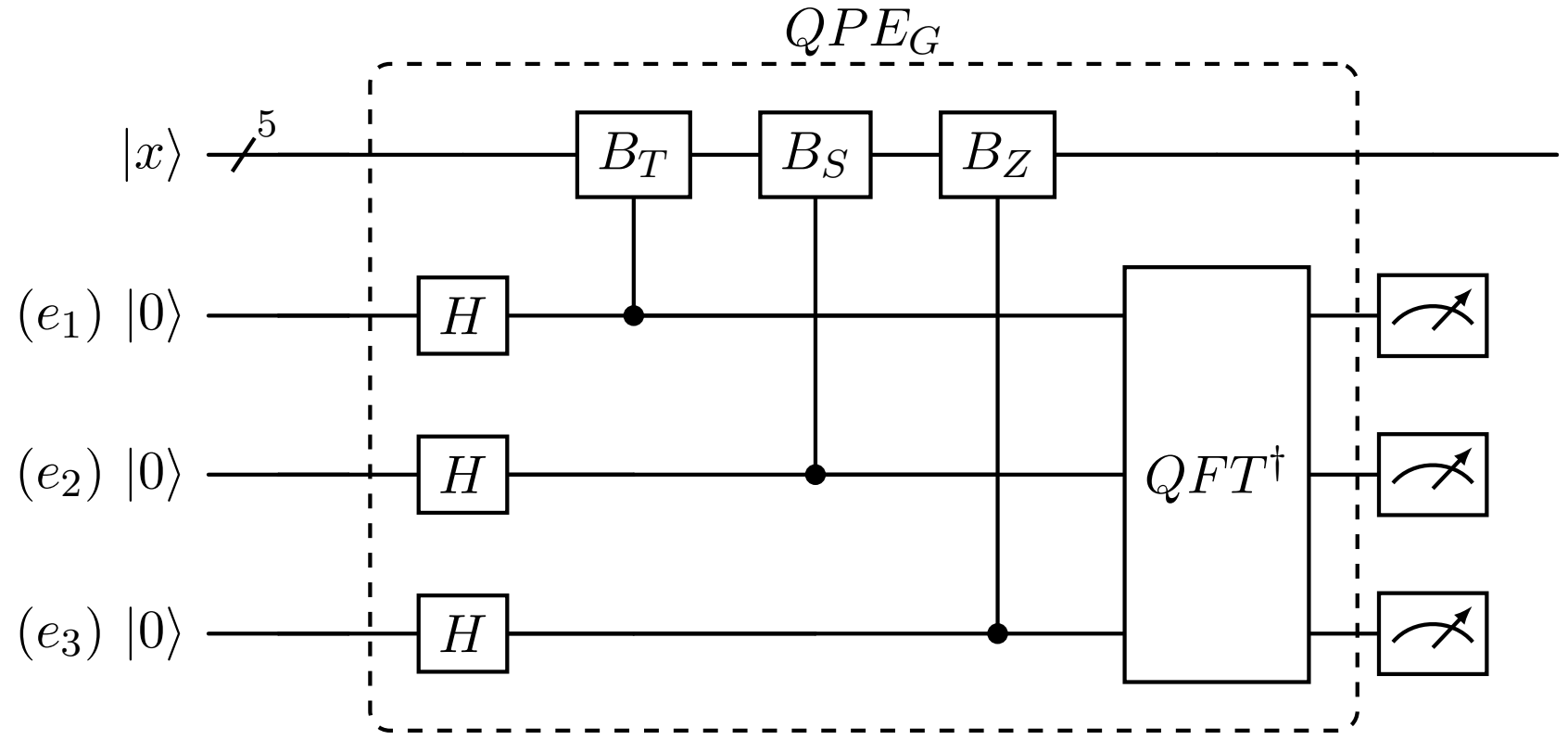
\begin{figure}[htbp]
\centering
\begin{quantikz}
\lstick{ $\ket{x}$} & \qwbundle{5} & \qw & \gategroup[4,steps=5,style={dashed,rounded
corners,inner sep=6pt}]{$QPE_G$}&   \gate{B_T} & \gate{B_S} & \gate{B_Z} & \qw & \qw & \qw \\
\lstick{($e_1$) $\ket{0}$ }&  & \qw & \gate{H} & \ctrl{-1} & \qw & \qw & \gate[wires=3]{QFT^\dagger} & \meter{} \\
\lstick{($e_2$) $\ket{0}$ }&  & \qw& \gate{H} & \qw & \ctrl{-2} & \qw & & \meter{} \\
\lstick{($e_3$) $\ket{0}$ }&  & \qw& \gate{H} & \qw & \qw & \ctrl{-3} & & \meter{}
\end{quantikz}
\caption{Internal structure of the $QPE_{G}$ block for the worked example ($m=3$ estimation qubits, $\theta=\pi/4$). Qubit $e_1$ holds the least significant bit of the estimated phase. The final measurements are shown only to discuss the quantum spectrum (Table~\ref{tab:qspectrum}); they are not part of the oracle}
\label{fig:qpe-block}
\end{figure}

Feeding the GEO into the phase-estimation construction of Theorem~\ref{thm:qpe-count} would require $m=\lceil\log_2| E(G)| \rceil + 1 = 4$ estimation qubits; however since $| E(G) |=5$ is not a power of $2$ by the remark following Theorem~\ref{thm:qpe-count} we can use $m=\lceil\log_2| E(G)|\rceil = 3$. Therefore, we set the value $\theta=2\pi/2^3=\pi/4$. Since $P(\theta)^k = P(k\theta)$, if we denote $B = U_{G}^{\pi/4}$ then the controlled powers of $B$ needed by the estimation qubits register are simply $B^{2^0}=B_T$, $B^{2^1}=B_S$, $B^{2^2}=B_Z$, where $B_T$, $B_S$ and $B_Z$ denote the same five-edge pattern of Figure~\ref{fig:encoding}, repeated with the standard gates $T$, $S$, and $Z$ in place of $P(\theta)$. Figure~\ref{fig:qpe-block} shows the full block.

If we fed to the $QPE_G$ instead of a single eigenvector $\ket{x}$ a superposition of all eigenvectors $2^{-N/2} \sum_{x=0}^{2^N-1} \ket{x}$ and were the three meters of Figure~\ref{fig:qpe-block} actually to be read out, the resulting probability distribution over $e_1e_2e_3$ would be exactly the \emph{quantum spectrum} of $G$ in the sense of~\cite{PelayoQPE2024}: the phase register reads $|E(G_S)|\in\{0,\dots,5\}$ with probability $2^{-N}T_{G}[|E(G_S)|]$, where $T_{G}[|E(G_S)|]$ is the number of subgraphs $G_S$ of $G$, with $S\subseteq V(G)$, having $|E(G_S)|$ edges (Definition~3.2 of ~\cite{PelayoQPE2024}). For our worked example, exhaustively classifying all $2^5=32$ subsets of $V(G)$ gives the spectrum of Table~\ref{tab:qspectrum}.

\begin{table}[htbp]
\centering
\begin{tabular}{ccc}
\toprule
$|E(G_S)|$ & Probability \% & $T_{G}[|E(G_S)|]$ \\
\midrule
0 & 37.50 & 12 \\
1 & 28.13 & 9 \\
2 & 18.75 & 6 \\
3 & 6.25  & 2 \\
4 & 6.25  & 2 \\
5 & 3.13  & 1 \\
\bottomrule
\end{tabular}
\caption{Quantum spectrum of $G$ for the worked example: probability of each outcome of the phase register in Figure~\ref{fig:qpe-block}, and the corresponding raw subgraphs count $T_{ G}[|E(G_S)|]$ that is the number of subsets $S \subseteq V(G)$ which induced subgraph has exactly $|E(G_S)|$ edges, following the convention of~\cite{PelayoQPE2024}.
}
\label{tab:qspectrum}
\end{table}

As an oracle, however, the phase register in Figure~\ref{fig:qpe-block} is never actually measured because doing so would collapse the superposition over $S$ and destroy the very interference Grover amplification relies on. 

\section{Building the clique oracle}\label{sec:counting-to-marking}

We are now in position to build the Grover oracle. As we did in the earlier worked example, and to simplify the discussion, we swap the roles of the graphs: we treat $\bar G$ as the original graph and $G$ as its complement, while keeping $G$ as the basis of our analysis.
For those state vectors $\ket{x}$ of the GEO register for which in the phase register all qubits are zero then we know that $S(x)$ is an independent set of $ G$ which, in turns, corresponds to a clique of $\bar G$. All it remains to do is to check that $|S(x)|$ is greater or equal than $k$. This can be done by using a new register, called \textit{weight} register, containing $n=\lceil \log_2 (N)\rceil+1$ qubits. 
The weight register is pre-loaded with the value of $\ket{-k}$ represented in two's complement. Then a set of $+1$ circuits controlled by the qubits of the GEO register are added to the circuit. In this way the most significant qubit (MSB) of the weight register will be $\ket{0}$ if and only if the state vector $\ket{x}$ of the GEO register satisfies $|S(x)|-k \geq 0$. 

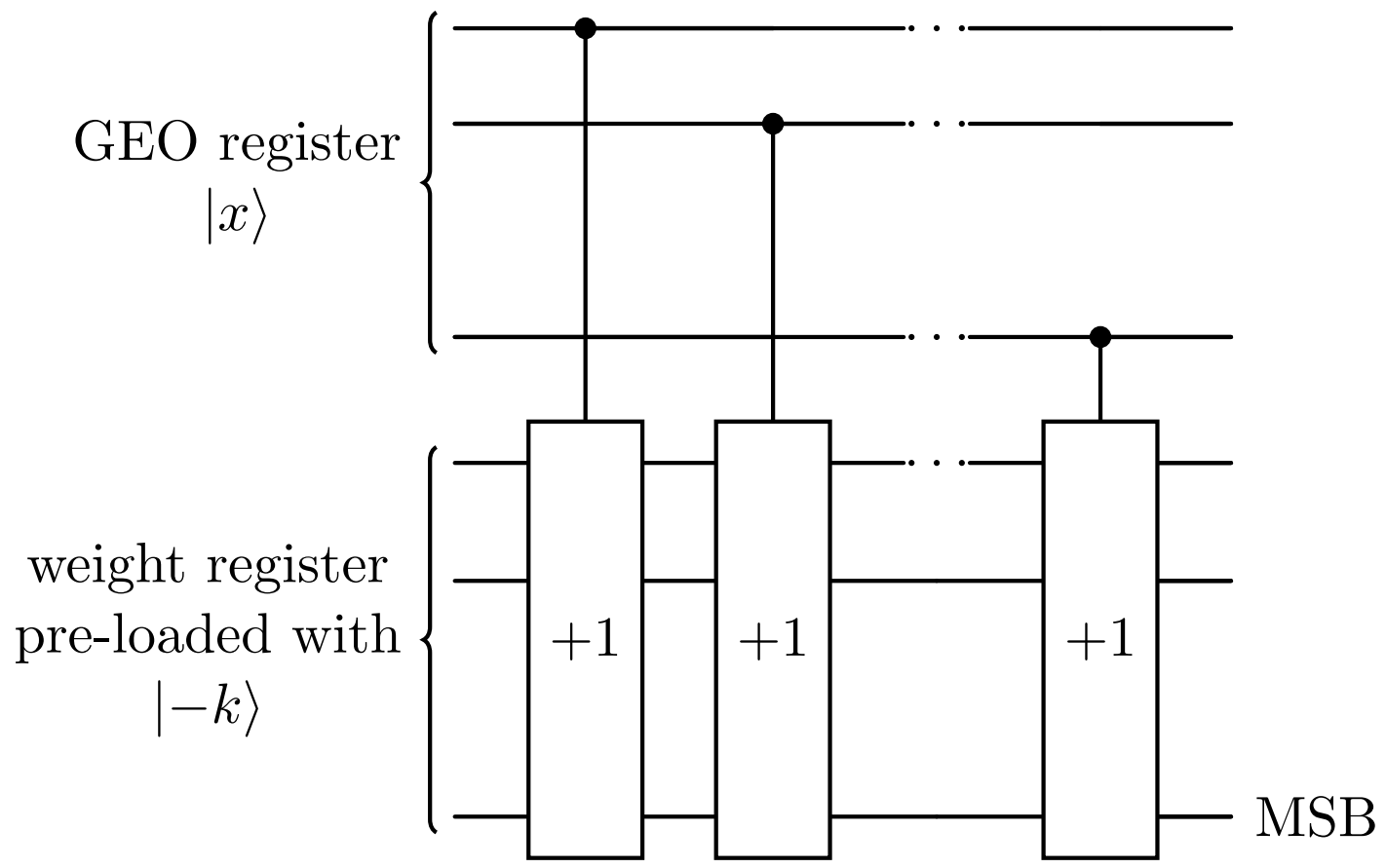
\begin{figure}[htbp]
\centering
\begin{quantikz}
\lstick[3]{GEO register\\ $\ket{x}$} &  \ctrl{5} & \qw & \ldots & \qw &  \qw  \\
 & \qw & \ctrl{4} & \ldots \qw &  & \qw  \\[.8cm]
& \qw & \qw &\ldots &  \ctrl{1} & \qw \\
\lstick[3]{weight register \\pre-loaded with \\$\ket{-k}$} & \gate[wires=3]{+1} & \gate[wires=3]{+1} & \ldots &   \gate[wires=3]{+1} & \qw \\
 & & \ldots  & & &  \qw \\[.8cm]
 & & \ldots & & &  \qw \rstick{\text{MSB}} 
\end{quantikz}
\caption{The circuitry we use to check if the state vector $\ket{x}$ of the GEO register satisfies $|S(x)|-k \geq 0$. We use one controlled increment per GEO qubit, all acting on the same qubits of the weight  register pre-loaded with $-k$ in two's complement. The MSB is zero if and only if $|S(x)|-k \geq 0$  }
\label{fig:weight-abstract}
\end{figure}
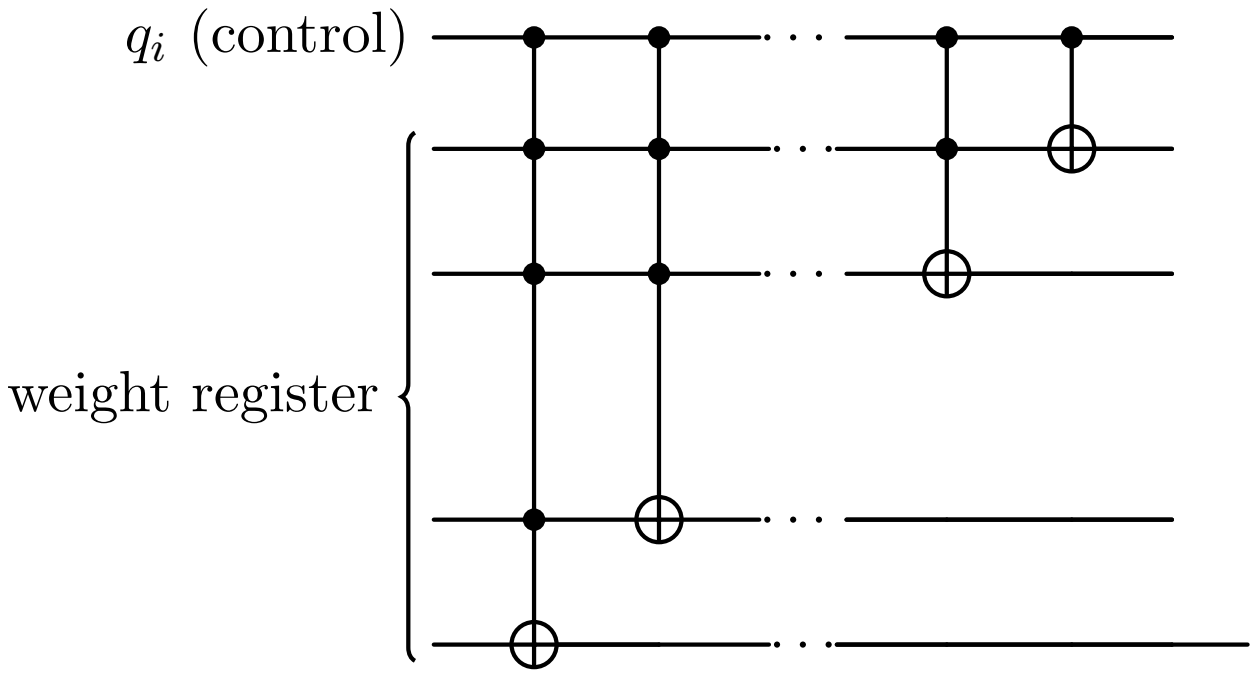
\begin{figure}[htbp]
\centering
\begin{quantikz}
\lstick{$q_i$ (control)} & \ctrl{1} & \ctrl{1} & \ldots\ & \ctrl{1} & \ctrl{1} & \qw \\
\lstick[4]{weight register} & \ctrl{1} & \ctrl{1} & \ldots &  \ctrl{1} & \targ{} & \qw \\
 & \ctrl{1} & \ctrl{1} & \ldots\ & \targ{} & \qw & \qw \\[0.8cm]
 & \ctrl{1} &  \targ{} & \ldots\ & \qw & \qw & \qw \\
 & \targ{} & \qw & \ldots & \qw & \qw & \qw &  
\end{quantikz}
\caption{Elementary decomposition of one controlled $+1$ block from Figure~\ref{fig:weight-abstract} (control $q_i$, target weight register): a sort of ripple-carry increment built from a (multi)-controlled NOT gate.}
\label{fig:weight-decomp}
\end{figure}

\noindent
Each controlled $+1$ block is a completely standard reversible ripple-carry incrementer, built from a cascade of multi-controlled NOT gates of decreasing arity (Figure~\ref{fig:weight-decomp}). 

Now we are able to complete our oracle. In what follows, we will describe only the \textit{computational} part of the oracle circuit, leaving it to the reader to derive the corresponding \textit{uncomputation} phase --which simply requires reversing the sequence of gates-- necessary to ensure the correctness of Grover's iteration.

\begin{figure}[htbp]
\centering
\begin{quantikz}
\lstick[3]{phase reg. \\ $\ket{0}^{\otimes m}$}   
& \gate{H} & \ctrl{3} & \qw  & \qw \cdots & \qw & \gate[3]{QFT^\dagger} & \qw & \qw &  \qw \cdots & \qw & \ctrl[open]{9} & \qw \rstick[3]{trigger \\if $\ket{0}^{\otimes m}$}\\
& \gate{H}        & \qw & \ctrl{3}   & \qw \cdots & \qw & \qw & & \qw & \qw \cdots & \qw & \ctrl[open]{8} & \qw \\[.98cm]
& \gate{H} & \qw & \qw & \qw \cdots & \ctrl{3} & \qw & & \qw & \qw \cdots & \qw & \ctrl[open]{7} & \qw \\
\lstick[3]{GEO reg. \\ $\ket{x}$}
& & \gate[3]{B^{2^0}} & \gate[3]{B^{2^1}} & \qw \cdots & \gate[3]{B^{2^{m-1}}} &  \qw & \ctrl{3} & \qw  & \qw \cdots& \qw & \qw & \qw \\
&  &                   &                   & \qw \cdots &                   & \qw &  \qw  & \ctrl{3}  & \qw \cdots & \qw & \qw & \qw \\
&  &                   &                   & \qw \cdots &                   & \qw & \qw &  & \qw \cdots & \ctrl{3} & \qw & \qw \\
\lstick[3]{weight reg. \\ preloaded \\with $\ket{-k}$}
 & \qw & \qw & \qw & \qw \cdots & & \qw & \gate[3]{+1} & \gate[3]{+1} & \qw \cdots & \gate[3]{+1} & \qw & \qw \\
& \qw & \qw & \qw & \qw \cdots & &\qw &             & \qw  & \qw \cdots &  & \qw & \qw \\[0.98cm]
& \push{\text{MSB}} & \qw & \qw & \qw \cdots & &\qw &             &             & \qw \cdots &             & \ctrl[open]{1} & \qw \rstick{trigger\\ if $\ket{0}$} \\
\lstick{flag $\ket{1}$}
& \gate{H} & \qw & \qw & \qw \cdots & \qw & \qw & \qw & \qw & \qw \cdots  & \qw & \targ{} & \qw
\end{quantikz}
\caption{The complete detailed construction of the oracle where $B=U_G^{\theta}$.}
\label{fig:clique-oracle}
\end{figure}
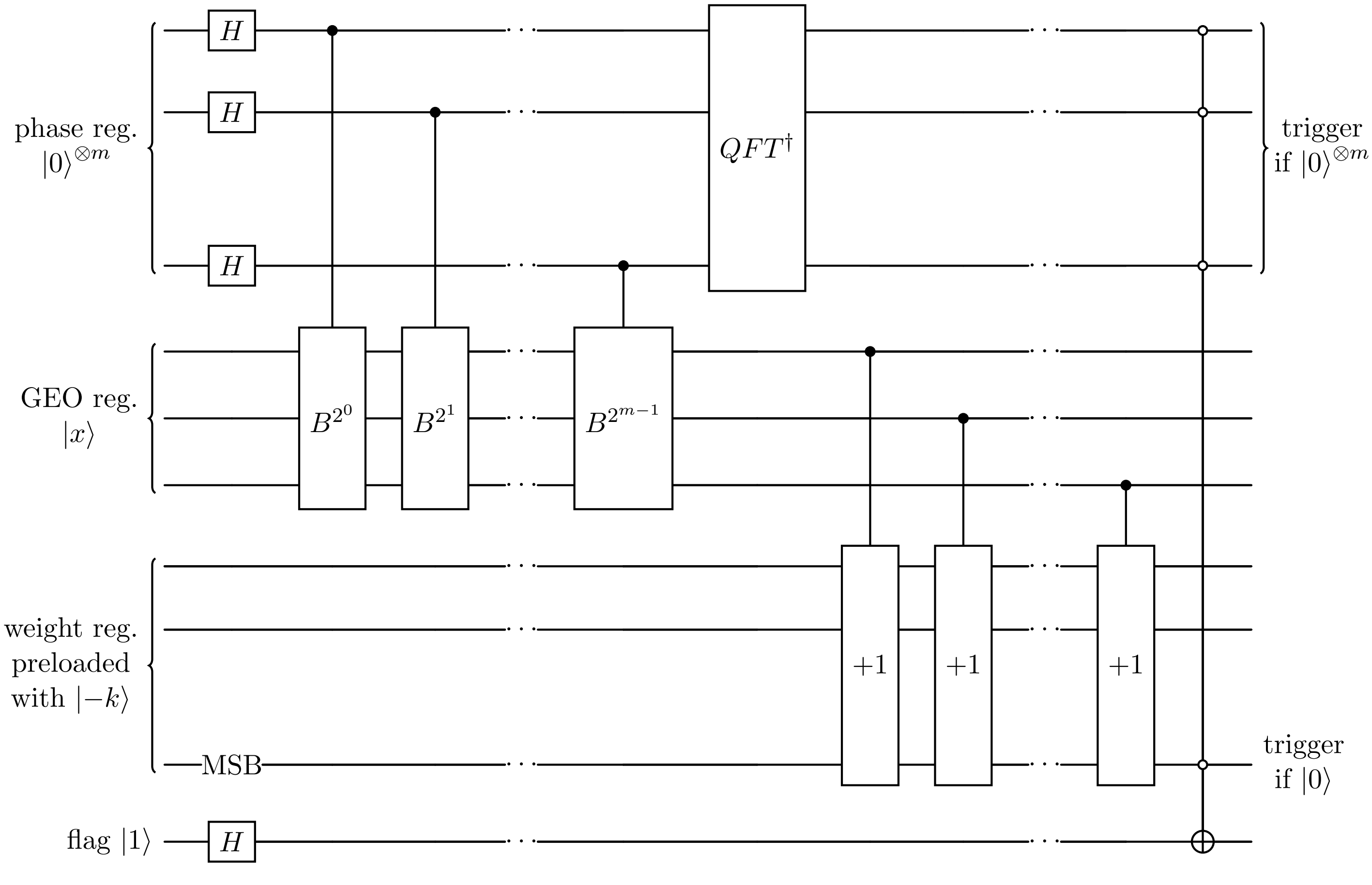

We have the phase register, the GEO register, the weight register and the flag. The complete circuit is reported in Figure \ref{fig:clique-oracle}. The phase register ends up to contain the phase of the eigenvector $\ket{x}$ of the GEO register, that, is as stated in Theorem \ref{thm:qpe-count}, it contains $|E(G_{S(x)})|$. If $|E(G_{S(x)})|$ and the MSB of the weight register are zero then we get the \textit{phase flip} required in the Grover iteration. 

\subsection{Correctness and complexity}\label{sec:correctness}

\begin{figure}[htbp]
\centering
\begin{quantikz}
\lstick{$\ket{0}^{\otimes m}$} &  \qwbundle{m} & \gate[2]{QPE_G} & \qw & \qw &  \qw & \qw & \qw & \ctrl[open]{5} & \qw &  \qw\\
\lstick{$\ket{x}$}             &  \qwbundle{N} & & \qw &\qw & \gate[4]{\text{weight}}  &  \qw& \qw &\qw & \qw &  \qw\\
\lstick[3]{$\ket{-k}$}       & \qw & \qw & \qw & & \qw & \qw & \qw& \qw & \qw &  \qw\\
                                & \qw & \qw & \qw & & \qw & \qw& \qw & \qw & \qw &  \qw \\[.8cm]
                               & \qw & \qw & \qw & \qw & \qw&  \qw & &  \ctrl[open]{1} &  \qw &  \qw \rstick{MSB}\\
\lstick{$\ket{1}$}             & \gate{H} & \qw & \slice{$\ket{\varphi_1}$}  & \qw & \qw & \slice{$\ket{\varphi_2}$} & \qw &  \targ{} & \slice{$\ket{\varphi_3}$} & \qw
\end{quantikz}
\caption{Schematic circuit of the oracle of Figure \ref{fig:clique-oracle}.}
\label{fig:clique-oracle-correctness}
\end{figure}
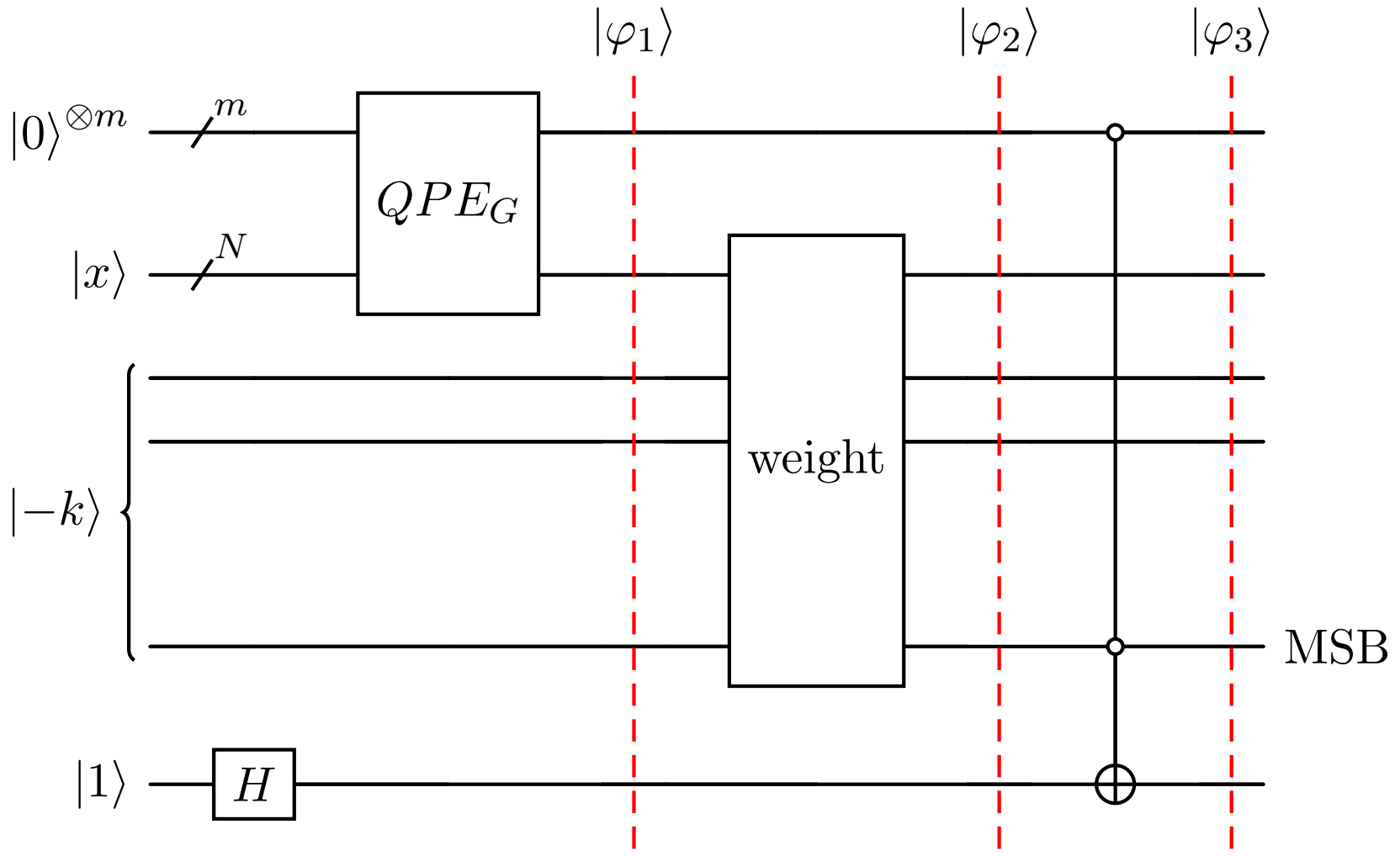
With reference to the schematic circuit of Figure \ref{fig:clique-oracle-correctness} now we review the behavior of the circuit. After the $QPE_G$ block the state of the circuit is, by Theorem \ref{thm:qpe-count}, the following:

\begin{equation}
    \ket{\varphi_1} = \ket{\ |E(G_{S(x)})| \ } \otimes \ket{x} \otimes  \ket{-k} \otimes  \ket{ -  }
\end{equation}
then, after the weight block, we obtain that $\ket{\varphi_2}$ is
\begin{equation}
    \ket{\varphi_2} = \ket{\ |E(G_{S(x)})| \ } \otimes \ket{x} \otimes  \ket{\ |S(x)|-k \ } \otimes \ket{-}
\end{equation}
at this point if $S(x)$ is an independent set of $G$ --and thus a clique of $\bar G$-- of cardinality at least $k$, we have that $|E(G_{S(x)})|=0$ and $|S(x)|-k\geq 0$ that is, the MSB of the weight register is zero. By using this information we can realize the phase flip. In fact if
\[
f(x)=
\begin{cases}
    1 & \text{ if $S(x)$ is a clique of cardinality $|S(x)|\geq k$}\\
    0 & \text{otherwise}
\end{cases}
\]
we have that $\ket{\varphi_3}$ is 
\begin{equation}
    \ket{\varphi_3} =-1^{f(x)} \ket{\ |E(G_{S(x)})| \ } \otimes \ket{x} \otimes \ket{\ |S(x)|-k \ } \otimes \ket{-}
\end{equation}
that is, we obtained the phase flip required for the Grover's oracle. 

\subsubsection{Complexity}\label{subsec:complexity}

Recall that we denote $N=|V(G)|$ and $M=|E(G)|$. As seen in the previous section the GEO register uses $N$ qubits, the phase register $\lceil\log_2 M \rceil+1$ qubits and the weight register uses $\lceil\log_2 N \rceil+1$ qubits. Therefore, since $M=O(N^2)$ we have  $O(N + \log N + \log N^2) = O(N)$ qubits in total.

As for the exponentiation of the GEO in the $QPE_G$ circuit we have the following. We note that the matrix expansion of the GEO $B=U_G^{\theta}$ is a diagonal matrix $diag(B)=(d_1, d_2,\dots, d_{2^N})$ whose elements are either $d_h=1$ or $d_h=e^{ia\theta}$ for some $a \neq 0$. Then the matrix $B^j$ is a diagonal matrix whose element are $diag(B^j)=(d_1^j, d_2^j,\dots, d_{2^N}^j)$. Therefore  we have that  either $d_h^j=1$ or $d_h^j=\left (e^{ia\theta} \right)^j=e^{i a (j \theta)}$. This means that $B^j=U_G^{j\theta}$. In other words, for $B^j$, we need the exact same circuit as  $B$, but with a phase angle of $j \theta$.

A single call to $QPE_{G}$ (forward or inverse) invokes the GEO a total of $O(\log_2 M)$ times (Section~4.3 of~\cite{PelayoQPE2024}) and each GEO contains $O(M)$ controlled phase gates; the weight circuit costs $O(N)$ reversible additions each of which contains $\log_2 N +1$ multi-controlled $X$ gate. The $QFT^\dagger$ require $O(m^2)=O(\log^2 M)$ gates. Hence each oracle call costs $O(N\log N+M \log M)$ elementary gates, and the full Grover search costs
\[
O\!\left(2^{N/2}\, (N \log N+ M \log M \right)
\]
gates in the worst case, i.e.\ the usual quadratic speed-up over the $O(2^N)$ classical exhaustive search --- Max-Clique remains NP-hard, and no super-polynomial speed-up is claimed or expected. As in~\cite{PelayoQPE2024}, we emphasize that this is a gate-count analysis and does not include the (separately hard) engineering cost of implementing the required phase-gate precision on real hardware.

\subsection{Numerical verification}\label{sec:verification}

We verified the oracle construction by implementing the full Grover algorithm and verified it on the worked example of Section~\ref{sec:worked-example}, checking every non-ancilla basis state after the complete compute--mark--uncompute sequence. 

We used first the graph of Figure \ref{fig:graph-example} which contains exactly one clique of cardinality 3. With threshold $k=3$, the oracle marked exactly this one state out of $32$, matching an exhaustive classical check; all $32/32$ phases matched the expected pattern and all ancilla registers (weight, estimation) returned exactly to their initial states. Table~\ref{tab:grover-results} reports the resulting Grover amplification after a single iteration, compared against the closed-form textbook probability $\sin^2\!\bigl((2\cdot 1+1)\theta\bigr)$ with $\theta=\arcsin\sqrt{1/32}$.

\begin{table}[htbp]
\centering
\begin{tabular}{lr}
\toprule
Metric & Value \\
\midrule
Probability before amplification & $0.0313$ \\
Probability after 1 iteration (simulated) & $0.2583$ \\
Probability after 1 iteration (closed form) & $0.2583$ \\
Amplification factor & $\times 8.27$ \\
\bottomrule
\end{tabular}
\caption{Grover amplification results for the worked example, simulated exactly and cross-checked against the standard closed-form Grover probability formula.}
\label{tab:grover-results}
\end{table}

The agreement between the simulated and closed-form probabilities (differing only in the third decimal place, attributable to floating-point rounding) confirms that the compute--mark--uncompute construction behaves as a textbook Grover oracle with no hidden side effects from the uncomputation steps.


\subsubsection{Scaling up: $N=5$ to $N=10$ on Qiskit}\label{sec:qiskit-verification}

As a further, fully independent check --- using a different simulation engine and shot-based sampling rather than exact linear algebra --- we implemented the complete Grover algorithm circuit 
in Qiskit and ran it on the \texttt{AerSimulator} across six graphs of increasing size, $N=5,\dots,10$, each with a single, maximum clique. According to what said in the previous sections the simulator acted on the complement $\bar G$ of those graphs to find the independent set of size $\geq k$. 
Table~\ref{tab:qiskit-graphs} describes each graph and the resulting qubit count and reports the probability of measuring the maximum clique after the theoretically optimal $r$ iterations, using a fixed budget of $1024$ shots for all $N$.

\begin{table}[htbp]
\centering
\begin{tabular}{clrrrrr}
\toprule
$N$ & $E(G)$ & $|E(\bar G)|$ & $k$ & Qubits & $r_{\text{opt}}$ &$p(\text{clique})$\\
\midrule
5  & $CQ(\{0,1,2\}) \cup \{\{0,3\}, \{2,4\}\}$ & 5  & 3 & 12 & 4 & 0.9990\\
6  & $CQ(\{0,1,\dots,3\}) \cup \{ \{0,4\},\{1,4\},\{2,5\},\{3,5\} \}$& 5 & 4 & 13 & 6 & 0.9963\\
7  & $CQ(\{0,1,\dots,3\}) \cup \{ \{0,4\},\{1,5\},\{2,6\} \}$ & 12  & 4 & 15 & 8 & 0.9956\\
8  & $CQ(\{0,1,\dots,4\}) \cup \{ \{0,5\},\{1,6\},\{2,7\} \}$ & 15 & 5 & 16 & 12 & 1.0000\\
9  & $CQ(\{0,1,\dots,5\}) \cup \{ \{0,6\},\{1,7\},\{2,8\} \}$ & 18 & 6 & 19 & 17 & 0.9995 \\
10 & $CQ(\{0,1,\dots,6\}) \cup \{ \{0,7\},\{1,8\},\{2,9\} \}$ & 21 & 7 & 20 & 25 & 0.9990\\
\bottomrule
\end{tabular}
\caption{The six worked graphs. With $CQ(S)$ we denoted the set of edges of the complete graph with vertex set $S$. Column Qubits is the total circuit size (graph $+$ weight $+$ estimation ancilla). $r_{\text{opt}}=\lfloor\frac{\pi}{4}\sqrt{2^N}\rfloor$ is the theoretically optimal number of Grover iterations when there is only one solution to the problem. The column $p(clique)$ is the probability of measuring the maximum clique, shot-based Qiskit/Aer execution, at the optimal $r_{\text{opt}}$, using $1024$ shots per instance. For $N=5,6$, values were cross-checked against exact state-vector computation and agreed to within expected sampling noise.}
\label{tab:qiskit-graphs}
\end{table}

In every one of the six cases the success probability reaches $99.5$--$100\%$ at $r_{\text{opt}}$, confirming convergence with no sign of degradation as $N$ grows. 

\section{The MHSP and the dualization of Boolean functions} \label{sec:MHSP_dualization}

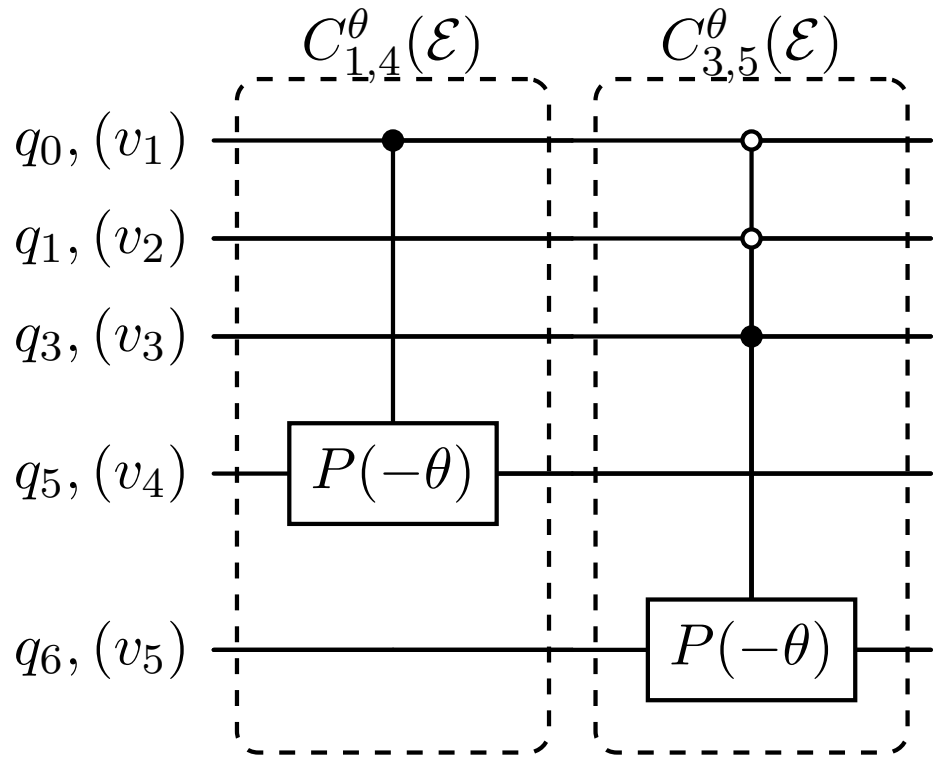
\begin{figure}
\centering
\begin{quantikz} 
\lstick{$q_0, (v_1)$} & \ctrl{3} \gategroup[5,steps=1,style={dashed,rounded
corners,inner sep=6pt}]{$C_{1,4}^{\theta}(\mathcal{E})$} & \qw       & \octrl{4}  \gategroup[5,steps=1,style={dashed,rounded
corners,inner sep=6pt}]{$C_{3,5}^{\theta}(\mathcal{E})$}& \qw  \\
\lstick{$q_1, (v_2)$} &  \qw     & \qw      & \octrl{3} & \qw   \\
\lstick{$q_3, (v_3)$} &  \qw     & \qw      & \ctrl{2} & \qw \\
\lstick{$q_5, (v_4)$} & \gate{P(-\theta)}    & \qw & \qw& \qw    \\
\lstick{$q_6, (v_5)$} & \qw   & \qw   &  \gate{P(-\theta)} & \qw
\end{quantikz}
    \caption{Two examples of multi-controlled phase gate $C_{j,h}^{\theta}(\mathcal{E})$ with $\mathcal{E}=\{0,1,3,5,6\}$: left $C_{1,4}^{\theta}(\mathcal{E})$, right $C_{3,5}^{\theta}(\mathcal{E})$.}
    \label{fig:multi-controlled-gateas}
\end{figure}

In this section, we are going to use the tools from the previous section in order to tackle the MHSP in hypergraphs. We will give a quantum circuit which is analogous to the graph encoding operator of~\cite{PelayoQPE2024} and was used in previous Sections\footnote{Code in Qiskit is available  \url{https://github.com/MauroMezzini/clique_and_hitting_set_oracles}}. This operator, if applied to any computational basis state $\ket{x}$,  gives us a state vector $\ket{\phi}= e^{i\varphi_x} \ket{x}$ in which the phase $\varphi_x$  of the element $\ket{x}$ encodes the number of hyperedges covered by $S(x)$. 
The circuit for the HEO will have $N=|V(\mathcal{H})|$ qubits. Its construction is given by Algorithm \ref{alg:hypergraph_phase_encoding}. Here is its description.
At each qubit $v$ we add the uncontrolled phase gate $P(|N_{\mathcal{H}}(v)| \theta)$. Then we add the multi-controlled phase gates defined as follows. 
\begin{definition}
Let $\mathcal H$ be a hypergraph, $\mathcal E \in \mathcal H$, $\mathcal{E} = \{v_1, v_2, ..., v_{|\mathcal{E}|}\}$ and $|\mathcal {E}|>1$. Given $v_j, v_h \in \mathcal E$ with $j<h$, the multi-controlled phase gate called $C_{j,h}^{\theta}(\mathcal{E})$ applies to qubit $v_h$, the phase gate $P(-\theta)$ controlled by all qubits $v_1,...,v_{j-1}$ when they are in state $\ket{0}$ and by the qubit $v_j$ when it is in state $\ket{1}$.
\end{definition}
\noindent
As an example, if $\mathcal{E}=\{0,1,3,5,6\}$ then $C_{1,h}^{\theta}(\mathcal{E})$ is the phase gate $P(-\theta)$ applied to $v_h \in \{1,3,5,6\}$ controlled by $v_1$ in state $\ket{1}$, while $C_{3,h}^{\theta}(\mathcal{E})$ is the phase gate $P(-\theta)$ applied to $v_h \in \{5,6\}$ controlled by $v_1$ and $v_2$ in state $\ket{0}$ and $v_3$ in state $\ket{1}$ (see Figure~\ref{fig:multi-controlled-gateas} for an example of $C_{1,4}^{\theta}(\mathcal{E})$ and $C_{3,5}^{\theta}(\mathcal{E})$). Note that  $C_{1,h}^{\theta}(\mathcal{E})$ is a controlled phase gate applied to qubit $v_h$ controlled by the single qubit $v_1$.\\

Now, for every hyperedge $\mathcal{E} \in E(\mathcal{H})$ such that $|\mathcal{E}|>1$, $\mathcal{E}=\{v_1,v_2,\dots, v_{|\mathcal{E}|}\}$, for every $v_j \in \mathcal{E}$ with $1 \leq j < |\mathcal{E}|$ and for every $h$ such that $j <h \leq |\mathcal{E}|$ we add the multi-controlled phase gate $C_{j,h}^{\theta}(\mathcal{E})$ to the circuit.

\begin{algorithm}
\caption{Quantum Circuit Construction for Hyperedge covering Encode}
\label{alg:hypergraph_phase_encoding}
\begin{algorithmic}[1]
\Require Hypergraph $\mathcal{H}$, phase angle parameter $\theta$
\Ensure Quantum circuit $D$ acting on $|V(\mathcal{H})|$ qubits 

\State Initialize an empty quantum circuit $D$ with $|V(\mathcal{H})|$ qubits
\Statex

\Comment{Step 1: Apply uncontrolled phase gates to each qubit}
\For{each qubit $v \in V(\mathcal{H})$}
    \State $h \leftarrow |N_{\mathcal{H}}(v)|$ \Comment{Number of hyperedges incident to $v$}
    \State Apply $P(h\theta)$ to qubit $v$ in circuit $D$
\EndFor
\Statex

\Comment{Step 2: Apply multi-controlled phase gates for hyperedges of size $> 1$}
\For{each hyperedge $\mathcal{E} \in E(\mathcal{H})$ with $|\mathcal{E}| > 1$} \label{line:hyperedge_selected}
    \State Order the vertices in $\mathcal{E}$ as $\{v_1, v_2, \dots, v_{|\mathcal{E}|}\}$
    \For{$j \leftarrow 1$ \textbf{to} $|\mathcal{E}| - 1$}
        \For{each target qubit $v_h \in \{v_{j+1}, \dots, v_{|\mathcal{E}|}\}$}
            \Statex \qquad \qquad \Comment{Apply gate $C_{j,h}^{\theta}(\mathcal{E})$ to qubit $v_h$:}
            \State Apply $P(-\theta)$ to target qubit $v_h$, controlled by:
            \Statex \qquad \qquad \qquad $\bullet$ Qubits $v_1, \dots, v_{j-1}$ in state $|0\rangle$
            \Statex \qquad \qquad \qquad $\bullet$ Qubit $v_j$ in state $|1\rangle$
        \EndFor
    \EndFor
\EndFor

\State \Return Circuit $D$
\end{algorithmic}
\end{algorithm}

\begin{example} \label{ex:hypergraph encoding}  
    Let ${\mathcal{H}} = \{\mathcal{E}_1, \mathcal{E}_2,\mathcal{E}_3,\mathcal{E}_4,\mathcal{E}_5\}$ where $\mathcal{E}_1=\{0,3\}, \mathcal{E}_2=\{0,4\}, \mathcal{E}_3=\{1,3, 4\}, \mathcal{E}_4=\{0,1, 2\}$ and $\mathcal{E}_5=\{2,3, 4\}$ (see Figure \ref{fig:hypergraph} for a bipartite representation of $\mathcal{H}$). We use $\theta =2 \pi/2^m$ with $m = 4$ in order to avoid negative angles in the phase, so that in this example $\theta =\pi/8$. Then $|N_{\mathcal{H}}(0)|=3$, $|N_{\mathcal{H}}(1)|=2$, $|N_{\mathcal{H}}(2)|=2$, $|N_{\mathcal{H}}(3)|=3$ and $|N_{\mathcal{H}}(4)|=3$. So the algorithm will add a phase gate to each qubit $v$ with an angle $\theta \times |N_{\mathcal{H}}(v)|$ (see Figure \ref{fig:hypergraph encoding}). 
    Next, we add the multi-controlled phase gate in which $\delta =-\pi/8$. In this example, there are 3 multi-controlled phase gates involving more than 2 qubits.
\end{example}

\begin{figure}
\centering
    \includegraphics[width=0.6 \textwidth ]{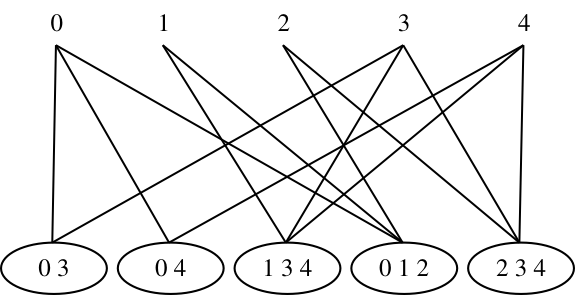}
    \caption{The hypergraph of Example \ref{ex:hypergraph encoding}, represented as a bipartite graph. The top row is the vertex set, and in the bottom row there are the hyperedges. }
    \label{fig:hypergraph}
\end{figure}
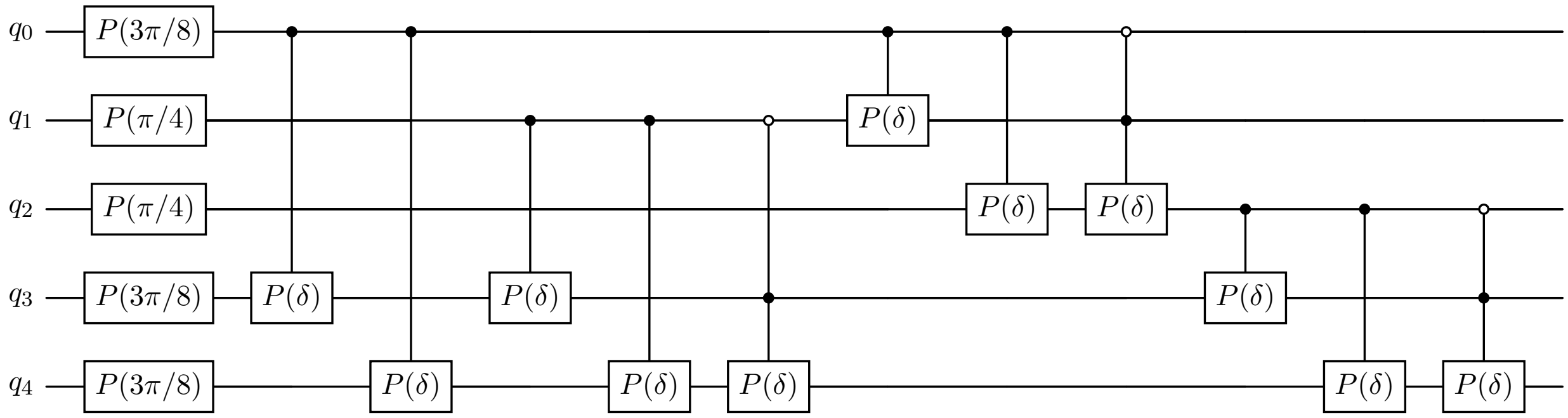
\begin{figure}
\resizebox{\textwidth}{!}{%
\begin{quantikz}
\lstick{$q_0$} &  \gate{P(3\pi/8)} & \ctrl{3} & \ctrl{4} & \qw       & \qw       & \qw       & \ctrl{1} & \ctrl{2} & \octrl{2} & \qw       & \qw       & \qw & \qw \\
\lstick{$q_1$} &  \gate{P(\pi/4)}  & \qw      & \qw      & \ctrl{2}  & \ctrl{3}  & \octrl{3} & \gate{P(\delta)} & \qw & \control{} & \qw       & \qw       & \qw & \qw \\
\lstick{$q_2$} &  \gate{P(\pi/4)}  & \qw      & \qw      & \qw       & \qw       & \qw       & \qw      & \gate{P(\delta)} & \gate{P(\delta)} & \ctrl{1}  & \ctrl{2}  & \octrl{2} & \qw \\
\lstick{$q_3$} &  \gate{P(3\pi/8)} & \gate{P(\delta)} & \qw & \gate{P(\delta)} & \qw & \control{1} & \qw & \qw    & \qw       & \gate{P(\delta)} & \qw & \control{} & \qw \\
\lstick{$q_4$} &  \gate{P(3\pi/8)} & \qw      & \gate{P(\delta)} & \qw & \gate{P(\delta)} & \gate{P(\delta)} & \qw & \qw & \qw & \qw & \gate{P(\delta)} & \gate{P(\delta)} & \qw
\end{quantikz}
}
    \caption{The circuit encoding the hypergraph of Example \ref{ex:hypergraph encoding}, in which $\delta = -\pi/8$.}
    \label{fig:hypergraph encoding}
\end{figure}
\begin{lemma} \label{le:multi-gate}
    Let $\mathcal H$ be a hypergraph and $\mathcal E \in \mathcal H$, $\mathcal E = \{v_1, v_2,\dots, v_{|\mathcal E|}\}$. If $\ket{x}$ is a computational basis state then  
    \begin{itemize}
        \item $C_{j,h}^{\theta}(\mathcal E)\ket{x}=e^{-i\theta}\ket x$ if and only if 
        \begin{enumerate}
            \item $v_j \in S(x) \cap \mathcal E$ and for all $v_z \in  S(x) \cap \mathcal E$, $j \leq z$ and
            \item $v_h \in S(x) \cap \mathcal E$
        \end{enumerate}
        \item $C_{j,h}^{\theta}(\mathcal E)\ket{x}=\ket x$ otherwise.
    \end{itemize}
\end{lemma}
\begin{proof}
    Let $\ket{x_0x_1\dots x_{N-1}}$ be the binary expansion of $\ket{x}$. By definition, $C_{j,h}^{\theta}(\mathcal E)$ applies the phase gate to qubit $\ket{x_{v_h}}$ if and only if qubits $\ket{x_{v_1}}=\ket{x_{v_2}}=\dots=\ket{x_{v_{j-1}}}=\ket{0}$ and $\ket{x_{v_j}}=\ket{1}$. Now, by contradiction, if  $v_j \notin S(x)$ then $\ket{x_{v_j}}=\ket{0}$ and the phase gate is not triggered. If, on other hand, $v_j \in S(x)$ but there exists a $v_z \in S(x) \cap \mathcal E$ such that $z < j$ then $\ket{x_{v_z}}=\ket{1}$ and the gate will not be triggered. This proves point (1). As for the point (2), if $v_h \notin S(x)\cap \mathcal E$ since by definition $v_h \in \mathcal E$ then we should have that $v_h \notin S(x)$, that is, $\ket{x_{v_h}}=  \ket{0}$. But by Lemma \ref{le:phase-gate}, the phase gate applied to a qubit in state $\ket{0}$ has no effect.
\end{proof}
\begin{lemma} \label{le:count-phase}
    Let $\mathcal H$ be a hypergraph and $\mathcal E \in \mathcal H$, $\mathcal E = \{v_1, v_2,\dots, v_{|\mathcal E|}\}$. If $\ket{x}$ is a computational basis state such that $|S(x) \cap \mathcal E|>1$ then there are exactly $|S(x) \cap \mathcal E|-1$ multi-controlled gates $C_{j,h}^{\theta}(\mathcal E)$ that will be triggered by $\ket{x}$.
\end{lemma}
\begin{proof}
Let $v_j \in S(x) \cap \mathcal E$ that satisfies point (1) of Lemma \ref{le:multi-gate}. Then, by point (2) of Lemma \ref{le:multi-gate}, for each $v_h \in ( S(x) \cap \mathcal E) \setminus \{v_j\}$ the multi-controlled gate $C_{j,h}^{\theta}(\mathcal E)$  will be triggered. Thus the Lemma follows. 
\end{proof}

\begin{theorem}
    Let $\mathcal{H}$ be a hypergraph and $D_{\mathcal{H}}$ be HEO in output to Algorithm \ref{alg:hypergraph_phase_encoding}. If  $\ket{x}$ is any computational basis vector then $D_{\mathcal{H}} \ket{x} =e^{i \varphi_x} \ket{x}$ and  $\varphi_x= |E_{S(x)}| \theta$.
\end{theorem}
\begin{proof}

Denote by $\ket{x_0x_1 \dots x_{N-1}}$ the binary expansion of $\ket{x}$. To determine the phase $\varphi_x$, we sum the contributions from Step 1 and Step 2 of the algorithm.
In Step 1, every qubit $v \in V(\mathcal{H})$ is acted upon by an uncontrolled phase gate $P(|N_{\mathcal{H}}(v)|\theta)$. By Lemma \ref{le:phase-gate}, qubit $v$ acquires this phase if and only if $\ket{x_v} = \ket{1}$ (i.e., $v \in S(x)$). Summing over all vertices in $S(x)$ gives:
\[
\varphi_x' = \sum_{v \in S(x)} |N_{\mathcal{H}}(v)| \theta.
\]
However, at this point, each hyperedge $\mathcal{E}$ covered by $S(x)$ is counted $|S(x) \cap \mathcal{E}|$ times in the phase $\varphi_x'$, which implies that a hyperedge is counted more than once if $|S(x) \cap \mathcal{E}|>1$.  Now, in Step 2 of the algorithm, consider a hyperedge $\mathcal{E} \in E(\mathcal{H})$ with $|\mathcal{E}| > 1$, ordered as $\{v_1, v_2, \dots, v_{|\mathcal{E}|}\}$ and consider all the multi-controlled gates $C_{j,h}^{\theta}(\mathcal{E})$ generated by Algorithm \ref{alg:hypergraph_phase_encoding} when the hyperedge $\mathcal{E}$ is selected at line \ref{line:hyperedge_selected}. If $|\mathcal{E} \cap S(x)| \leq 1$, none of these gates is triggered. Otherwise, by Lemma \ref{le:multi-gate} and Lemma \ref{le:count-phase}, there are precisely $|S(x) \cap \mathcal{E}|-1$ of such gates that are triggered. 
If we repeat this for every hyperedge in which $S(x) \cap \mathcal{E} \neq \emptyset$, that is for all hyperedges $\mathcal{E} \in E_{S(x)}$, we have that 
\begin{align*}
\varphi_x &= \varphi_x' -  \theta \sum_{\mathcal{E} \in E_{S(x)}} \left (|S(x) \cap \mathcal{E}|-1 \right) \\
& =\theta \sum_{\mathcal{E} \in E_{S(x)}} \left (|S(x) \cap \mathcal{E}| \right)-\theta \sum_{\mathcal{E} \in E_{S(x)}} \left (|S(x) \cap e|-1 \right) = |E_{S(x)}|\theta
\end{align*}
So we encode in the phase of $\ket{x}$ the number of hyperedges covered by $S(x)$.
\end{proof}
\begin{example}[continued from Example 1]
Consider the hypergraph of Example \ref{ex:hypergraph encoding} and let $\ket{x} = \ket{31} = \ket{11111}$ so that $S(x)=\{0,1,2,3,4\}$. The unconditional phase gate of Step 1 adds to $\ket{x}$ a phase $\varphi_x = 13 \theta$, which is exactly the number of edges of the bipartite graph of Figure \ref{fig:hypergraph}. Now only the gates $C_{1,2}^{\theta}(\mathcal{E}_1)$, $C_{1,2}^{\theta}(\mathcal{E}_2)$, $C_{1,2}^{\theta}(\mathcal{E}_3)$, $C_{1,3}^{\theta}(\mathcal{E}_3)$, $C_{1,2}^{\theta}(\mathcal{E}_4)$, $C_{1,3}^{\theta}(\mathcal{E}_4)$, $C_{1,2}^{\theta}(\mathcal{E}_5)$ and $C_{1,3}^{\theta}(\mathcal{E}_5)$ are  triggered. In the end, the total phase of $\ket{x}$ is $5\theta$ as all the hyperedges of $\mathcal{H}$ are covered by $S(x)$.
\end{example}

\begin{figure}[htbp]
\centering
\begin{quantikz}
\lstick{$\ket{x}$} & \qwbundle{N} & \qw      & \gate{HEO} & \phase{e^{-i|E(\mathcal{H})| \theta}} & \qw &\rstick{$e^{i(|E_{S(x)}|-|E(\mathcal{H})|)\theta}\ket{x}$}
\end{quantikz}
\caption{Schematic circuit of the HEO. Note that we add at the end of the circuit a global phase $e^{-i |E(\mathcal H)|\theta}$. In this way, if we measure the $QPE_{\mathcal {H}}$ register, we obtain $|E_{S(x)}|-|E(\mathcal{H})|$, which is zero if and only if $S(x)$ is a hitting set of $\mathcal H$.}
\label{fig:hitting-set-qpe}
\end{figure}
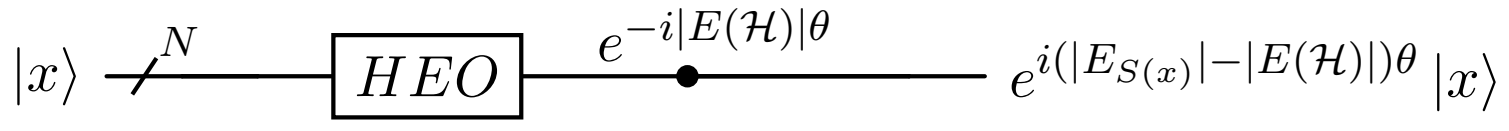

Since, by definition, a hitting set intersects with all hyperedges of the hypergraph, if for an $x$ the number of hyperedges covered by $S(x)$ is equal to the number of hyperedges of the hypergraph, then $S(x)$ is a hitting set. So if we apply at the end of the HEO circuit a global phase $e^{-i |E(\mathcal H)|\theta}$, we have that $S(x)$ is a hitting set if and only if $\varphi_x=0$ (see Figure \ref{fig:hitting-set-qpe}).

\subsection{Oracle construction}
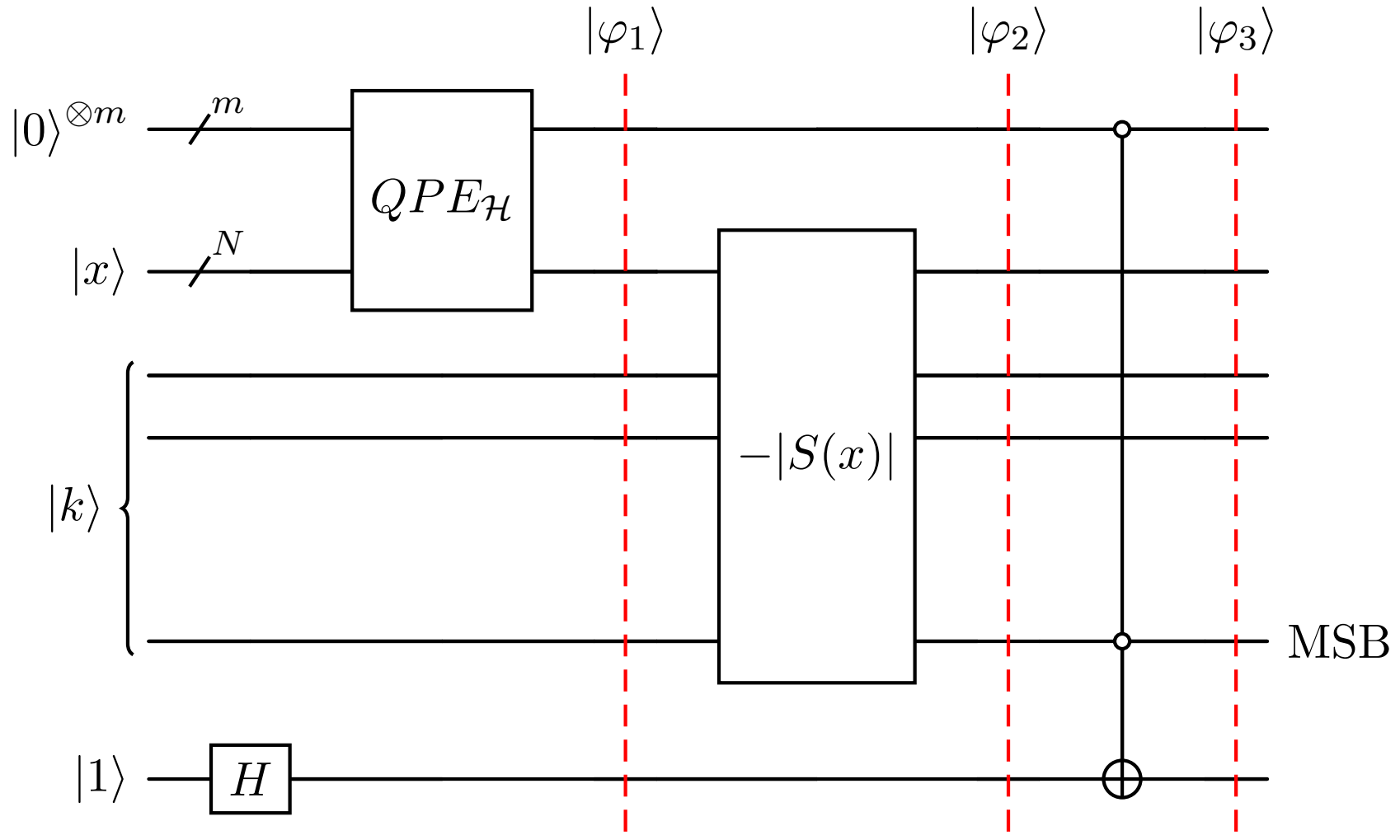
\begin{figure}[htbp]
\centering
\begin{quantikz}
\lstick{$\ket{0}^{\otimes m}$} &  \qwbundle{m} & \gate[2]{QPE_{\mathcal{H}}} & \qw & \qw &  &  \qw & \qw & \ctrl[open]{5} & \qw &  \qw  \\
\lstick{$\ket{x}$}             &  \qwbundle{N} &  \qw &\qw & \qw & \gate[4]{ -|S(x)|}  &  \qw& \qw &\qw & \qw &  \qw  \\
\lstick[3]{$\ket{k}$}       & \qw & \qw & \qw & \qw & \qw & \qw & \qw& \qw & \qw &  \qw \\
                                & \qw & \qw & & \qw & \qw& \qw & \qw & \qw &  \qw &  \qw \\[.8cm]
                               & \qw & \qw & \qw & \qw & \qw&  \qw & &  \ctrl[open]{1} &  \qw &  \qw \rstick{MSB}\\
\lstick{$\ket{1}$}             & \gate{H} & \qw &\slice{$\ket{\varphi_1}$} & \qw  & \qw &  \slice{$\ket{\varphi_2}$} & \qw &   \targ{} & \slice{$\ket{\varphi_3}$} & \qw
\end{quantikz}
\caption{Schematic circuit of the MHSP oracle. The only differences with the clique oracle (Figure \ref{fig:clique-oracle-correctness}) are that the weight register is pre-loaded with $k$ and instead of add one controlled circuit we used the controlled subtract-one circuit.}
\label{fig:hitting-set-oracle-correctness}
\end{figure}

We can use the HEO and the Quantum Phase Estimation algorithm to obtain, in the phase register, the value of $|E_{S(x)}|-|E(\mathcal{H})|$. The oracle's circuit will use the HEO register in the same way as the GEO is used in the clique oracle (see Section \ref{sec:counting-to-marking}). Since in the MHSP we want to check whether a subset of vertices is a hitting set of cardinality less than or equal to $k$, we pre-load the weight register with $\ket{k}$ and instead of using a controlled add-one circuit we will use a subtract-one circuit controlled by the qubits of the HEO register. In this way, if $k- |S(x)|<0$ then the MSB of the weight register is set to one, thus signalling that the set $S(x)$ has cardinality greater than $k$. 
Finally, using the phase register and MSB of the weight register we can make the phase flip required for the Grover's oracle. In fact, with reference to Figure~\ref{fig:hitting-set-oracle-correctness} 
\begin{equation}
    \ket{\varphi_1} = \ket{\ |E_{S(x)}|- |E(\mathcal {H})|\ } \otimes \ket{x} \otimes  \ket{k} \otimes  \ket{ -  }
\end{equation}
Then, after the weight block (labeled as $-|S(x)|)$, we obtain that $\ket{\varphi_2}$ is
\begin{equation}
    \ket{\varphi_2} = \ket{\ |E_{S(x)}|- |E(\mathcal {H})| \ } \otimes \ket{x} \otimes  \ket{\ k-|S(x)| \ } \otimes \ket{-}
\end{equation}
At this point, if $S(x)$ is a hitting set of $\mathcal{H}$ of cardinality at most $k$, we have that  the MSB of the weight register is zero. Therefore if
\[
f(x)=
\begin{cases}
    1 & \text{ if $S(x)$ is a hitting set of cardinality $|S(x)|\leq k$}\\
    0 & \text{otherwise}
\end{cases}
\]
we have that $\ket{\varphi_3}$ is 
\begin{equation}
    \ket{\varphi_3} =-1^{f(x)} \ket{\ |E_{S(x)}|- |E(\mathcal {H})| \ } \otimes \ket{x} \otimes \ket{\ k- |S(x)| \ } \otimes \ket{-}
\end{equation}
That is, we obtained the phase flip required for Grover's oracle. 

\subsection{Dualization of positive Boolean functions}
We can also, by measuring the $QPE_{\mathcal{H}}$ register, estimate the number of hitting sets of a hypergraph, and this is already an interesting result in itself. 
In fact, in the dualization problem of positive boolean functions (see \cite{Mezzini2024}) represented in a positive irredundant disjunctive normal form (PIDNF), we have that just counting the number of hitting sets of the corresponding hypergraph solves the dualization problem \cite{Mezzini2024}. In fact, by (\cite{Mezzini2024} Lemma 4), a PIDNF having $N$ Boolean variables is self-dual if and only if, in the corresponding hypergraph, the number of hitting sets is equal to $2^{N-1}$.
We can obtain an approximation of the number of hitting sets of the hypergraph by setting the $QPE_{\mathcal H}$ circuit and giving as input instead of $\ket{x}$ the superposition of all eigenvectors $2^{-N/2}\sum_{x=0}^{2^N-1} \ket{x}$. If we measure the frequencies in the phase register as we did for the corresponding $QPE_G$ of Figure~\ref{fig:qpe-block}, we can obtain an approximation of the number of hitting sets of $\mathcal H$ by measuring the frequencies of phase $|E_{S(x)}|-|E(\mathcal{H})|$ when $|E_{S(x)}|-|E(\mathcal{H})|=0$.
For the hypergraph of Example~\ref{ex:hypergraph encoding}, we may obtain a table similar to the quantum spectrum of a graph of Table \ref{tab:qspectrum}. In this case, we measure the number of subsets $S \subseteq V(\mathcal {H})$ that cover exactly $|E_{S}|$ hyperedges.

\begin{table}[htbp]
\centering
\begin{tabular}{ccc}
\toprule
$|E_{S(x)}|$ & Probability \% & \#(subsets of $V(\mathcal H)$)  \\
\midrule
0 & 3.13 & 1 \\
1 & 0 & 0 \\
2 & 6.25 & 2 \\
3 & 12.50  & 4 \\
4 & 28.13  & 9 \\
5 & 50.00  & 16 \\
\bottomrule
\end{tabular}
\caption{The probability of each outcome of the phase register for the hypergraph of Figure \ref{fig:hypergraph} and the corresponding raw vertex subset count. Since the hypergraph of the example is self-dual, we have that the number of hitting sets, that is, the number of subsets of $V(\mathcal H)$ which cover all the five hyperedges of the hypergraph is exactly $2^{N-1}$.}
\label{tab:hypergraph_spectrum}
\end{table}

\subsection{Complexity}

The HEO circuit requires $O(N)$ qubits and each hyperedge $\mathcal E$ requires $O(|\mathcal E|^2)=O(N^2)$ multi-controlled phase gates. Therefore, the total number of gates are $O(MN^2)$. As in the case of clique oracle, the $QPE_{\mathcal H}$ will require $O( \log M )$ additional qubits and the weight register $O(\log N)$ qubits. Thus the complete oracle requires $O(N+\log M+ \log N)$ qubits. This means that even if $M=O(2^N)$, that is, the hypergraph contains an exponential number of hyperedges, the number of qubits required is $O(N+ \log 2^N)=O(N)$.
As for the clique oracle, if $D_{\mathcal{H}}$ is the HEO in output from Algorithm \ref{alg:hypergraph_phase_encoding}, the exponentiation $D_{\mathcal{H}}^j$  can be made by simply using the same algorithm but passing to it the phase parameter $j \theta$, since the matrix expansion of $D_{\mathcal{H}}$ is a diagonal matrix $diag(d_1,\dots, d_{2^N})$ whose elements are of the form $d_h= e^{ia\theta}$ for some $a$. 
Since the oracle for the HEO is called $\log_2 M$ times, the total gate cost of the oracle is $O(N \log N+ MN^2 \log M)=O(MN^2 \log M)$ which is polynomial in the dimension of the hypergraph.

\section{Discussion and related work}\label{sec:discussion}

\subsection{Related work}\label{sec:related}

Quantum approaches to the MCP and its complementary formulations have been studied mainly within two frameworks: adiabatic quantum computing or QUBO formulations (see, e.g.,~\cite{ChildsFarhiGoldstoneGutmann2000,CaludeDinneenHua2017TCS,PelofskeHahnDjidjev2019,HuaDinneen2020ImprovedQUBO}) and gate-based quantum circuits. Within the gate-based model, besides variational algorithms such as QAOA and VQE~\cite{ChatterjeeMohtashimKundu21,Krpan2025}, continuous-time quantum walks~\cite{LiCTQW2021,LiSeidel2024}, and photonic implementations~\cite{YinPhotonic2023}, the main line of research relies on building an explicit oracle circuit and amplifying valid solutions through Grover's search~\cite{Grover96} or quantum minimum-finding routines~\cite{DurrHoyer1996}. A comprehensive overview of both classical and quantum methods for the MCP can be found in the recent survey by Marino et al.~\cite{Marino2026}. Here we focus our review on gate-based algorithms that construct Grover oracles, before discussing extensions to hypergraphs.

One of the earliest explicit Grover-oracle constructions for the MCP was proposed by Boji\'c~\cite{Bojic2012}, where vertex subsets are encoded in a register of $N$ qubits and pairwise adjacency is verified through a cascade of multi-controlled operations. Following a similar paradigm, Chang et al.~\cite{ChangEtAl2018} presented a circuit implementation with a detailed query-complexity analysis, comparing its asymptotic speed-up against classical branch-and-bound methods on the decision version of the problem. Later, Metwalli, Le Gall and Van Meter~\cite{MetwalliLeGallVanMeter2020} addressed the fixed-size $k$-clique and triangle-finding problems by preparing Dicke states so as to restrict the uniform superposition exclusively to subsets of cardinality $k$, with an experimental evaluation of circuit depth and noise on NISQ devices.

Several subsequent works focused on automating the generation of the oracle circuit and analyzing its exact gate and ancilla requirements. Sanyal et al.~\cite{SanyalBhaduriEtAl2022} developed a framework that constructs, for any input adjacency matrix and target size $k$, an oracle composed of multi-controlled Toffoli (MCT) edge-detector and clique-detector blocks, finding the maximum clique by iteratively decreasing $k$ from $N$. A later extension~\cite{SanyalBhaduriQudits2023} reduced the MCT decomposition overhead by employing higher-dimensional qudits for intermediate steps. Along similar lines, Dias and Zateska~\cite{DiasZateska2021} and Haverly and L\'opez~\cite{HaverlyLopez2021,Haverly2021Thesis} independently formulated Grover oracles for the MCP and derived explicit bounds on the number of qubits and elementary gates required.

A common feature of these algorithms is that, in order to find the clique number $\omega(G)$ from the decision oracle, Grover's search must be executed repeatedly for different values of the threshold $k$. Depending on whether $k$ is scanned linearly or via binary search, this requires $O(N)$ or $O(\log N)$ separate runs and measurements. To avoid these restarts, Han, Zheng, Chen and Wang~\cite{HanZhengChenWang2025} recently proposed a two-stage algorithm in which a classical pre-processing step based on Tur\'an's theorem bounds the feasible range of $\omega(G)$, and a dynamic quantum register updates the size of the largest clique found during the quantum search itself. This reduces the overall number of measurements to $O(1)$ while maintaining $O(\sqrt{2^N})$ Grover iterations. Nevertheless, like the previous circuit-model approaches, the underlying oracle in~\cite{HanZhengChenWang2025} still verifies edges explicitly through pairwise MCT gates, which requires $O(N^2)$ qubits in total.

When moving from graphs to hypergraphs, and specifically to the Minimum Hitting Set Problem (MHSP) and the dualization of positive Boolean functions~\cite{Mezzini2024}, existing quantum approaches have relied almost exclusively on QUBO and higher-order unconstrained binary optimization (HUBO) models for quantum annealers. In that setting, representing hyperedges of cardinality $|\mathcal{E}| > 2$ produces higher-order penalty terms that must be reduced to quadratic interactions by introducing auxiliary variables, which rapidly increases the number of physical qubits needed for minor embedding. On the other hand, if one attempts to generalize standard MCT-based gate-model oracles directly to a hypergraph $\mathcal{H}$ with $M = |E(\mathcal{H})|$ hyperedges, verifying whether a subset $S(x) \subseteq V(\mathcal{H})$ intersects every hyperedge typically requires allocating one ancilla qubit per hyperedge before applying a multi-controlled operation over all $M$ flags. While this requires $O(M)$ ancilla qubits in general, in the context of positive Boolean function dualization the number of hyperedges can be exponential in the number of vertices, i.e., $M \in O(2^N)$~\cite{Mezzini2024}, making any oracle that scales linearly with $M$ in space unfeasible.

\subsection{Comparison with existing circuit-model algorithms}\label{sec:comparison}

In contrast to prior MCP circuits that allocate dedicated ancilla qubits to check pairwise vertex adjacency, the approach presented in this work builds the Grover oracle from the GEO introduced in~\cite{PelayoQPE2024}. By running the phase estimation circuit $QPE_{\bar{G}}$ on the complement graph $\bar{G}$, the number of edges $|E(\bar{G}_{S(x)})|$ induced by each subset $S(x)$ is accumulated directly in the phase of the computational basis state $\ket{x}$ and read out coherently into the phase register, while the cardinality condition $|S(x)| \geq k$ is verified using a two's-complement weight register without additional comparator ancillas. The same principle extends to hypergraphs through the HEO of Algorithm~\ref{alg:hypergraph_phase_encoding}, where the overcounting of hyperedges intersecting $S(x)$ in multiple vertices is canceled directly in phase before applying $QPE_{\mathcal{H}}$.

In Table~\ref{tab:comparison} we summarize the computational complexity of our GEO and HEO oracles alongside existing Grover-based algorithms for the MCP, using the evaluation metrics of~\cite{HanZhengChenWang2025}.

\begin{table}[htbp]
\centering
\small
\resizebox{\textwidth}{!}{%
\begin{tabular}{lccccc}
\toprule
Method & Qubit cost & Oracle gate cost & Grover iterations & Measurements & Total gates \\
\midrule
Haverly and L\'opez~\cite{HaverlyLopez2021} & $O(N^2)$ & $O(\sqrt{N}\,2^N)$ & $O(N 2^N)$ & $O(N)$ & $O(kN^{N/2}2^{2N})$ \\
Sanyal et al.~\cite{SanyalBhaduriEtAl2022} & $O(N^2)$ & $O(\sqrt{N}\,2^N)$ & $O(N\sqrt{2^{N\log_2 N}})$ & $O(N)$ & $O(k\sqrt{N}\,2^{N+\frac{N\log_2 N}{2}})$ \\
Dias and Zateska~\cite{DiasZateska2021} & $O(N^2)$ & $O(N^2)$ & $O((N\log_2 N)^2\sqrt{2^N})$ & $O(\log_2 N)$ & $O(kN^2(\log_2 N)2^{N/2})$ \\
Chang et al.~\cite{ChangEtAl2018} & $O(N^2)$ & $O(N^2)$ & $O(N\sqrt{2^N})$ & $O(N)$ & $O(kN^22^{N/2})$ \\
Han et al.~\cite{HanZhengChenWang2025} & $O(N^2)$ & $O(N^2)$ & $O(\sqrt{2^N})$ & $O(1)$ & $O(N^2 2^{N/2} + N^{3/2}2^N)$ \\
\midrule
\textbf{This work (GEO -- MCP)} & $\mathbf{O(N)}$ & $O(|E(\bar{G})|\log|E(\bar{G})| + N\log N)$ & $O(\sqrt{2^N/M_k})$ & $O(\log N)^\dagger$ & $O\!\left(\log N \cdot 2^{N/2}(|E(\bar{G})|\log|E(\bar{G})| + N\log N)\right)^\dagger$ \\
\textbf{This work (HEO -- MHSP)} & $\mathbf{O(N + \log M)}$ & $O(M N^2 \log M + N\log N)$ & $O(\sqrt{2^N/M_k})$ & $O(\log N)^\dagger$ & $O\!\left(\log N \cdot 2^{N/2}(M N^2 \log M + N\log N)\right)^\dagger$ \\
\bottomrule
\end{tabular}%
}
\caption{Comparison with existing Grover-based circuit constructions for the MCP and extension to the MHSP on hypergraphs (figures for the first five rows as reported in~\cite{HanZhengChenWang2025}; $^\dagger$ figures for finding the optimum $\omega(G)$ or minimum hitting set size via binary search over $k$). Here $N=|V|$, $|E(\bar{G})|\le\binom{N}{2}$ is the number of edges in the complement graph $\bar{G}$, $M=|E(\mathcal{H})|\le 2^N$ is the number of hyperedges, and $M_k$ denotes the number of valid solutions for threshold $k$.}
\label{tab:comparison}
\end{table}

The main difference in Table~\ref{tab:comparison} lies in the space complexity. All previous gate-model constructions for the MCP require $O(N^2)$ qubits due to the ancilla registers needed to store intermediate edge checks. By contrast, our GEO oracle requires $N$ qubits for the vertex register, $\lceil \log_2 |E(\bar{G})| \rceil + 1$ qubits for the phase register, $\lceil \log_2 N \rceil + 1$ qubits for the weight register, and one flag qubit, yielding a total of $O(N)$ qubits. For hypergraphs, since the phase register only needs to represent values up to $M = |E(\mathcal{H})|$, the HEO oracle requires $O(N + \log M)$ qubits. As noted in Section~\ref{sec:MHSP_dualization}, this implies that even when the hypergraph has an exponential number of hyperedges, $M \in O(2^N)$, the total number of qubits remains bounded by $O(N + \log_2 2^N) = O(N)$, avoiding the exponential space blow-up of ancilla-per-hyperedge encodings.

This reduction in qubit count is traded against circuit depth and phase-gate operations. However, as shown in Section~\ref{subsec:complexity}, because the matrix representations of both $U_{\bar{G}}^\theta$ and $D_{\mathcal{H}}$ are diagonal in the computational basis, the controlled powers $(U_{\bar{G}}^\theta)^{2^j}$ and $D_{\mathcal{H}}^{2^j}$ required by Quantum Phase Estimation do not need to repeat the base circuit $2^j$ times; each power is implemented by the exact same circuit with a rescaled phase angle $2^j\theta$. Consequently, a single call to the oracle invokes the encoding operator only $m = O(\log M)$ times, resulting in $O(|E(\bar{G})|\log|E(\bar{G})| + N\log N)$ gates for the MCP and $O(MN^2\log M + N\log N)$ gates for the MHSP. Notice also that for the MCP the gate cost depends on $|E(\bar{G})|$, the number of edges of the complement graph, rather than being fixed at $\Theta(N^2)$. In practice, empirical studies on classical exact solvers (such as CliSAT~\cite{SanSegundo2023}) indicate that the hardest benchmark instances for maximum clique typically occur at high graph densities of $80\%$ to $95\%$~\cite{McCreesh2019,Marino2026}. In this dense regime for $G$, the complement $\bar{G}$ is sparse ($|E(\bar{G})| \ll \binom{N}{2}$), which is precisely when the GEO circuit requires the fewest controlled-phase gates.

Regarding the number of measurements needed to solve the optimization version of the problem, using a classical binary search over the threshold $k \in \{1,\dots,N\}$ requires $O(\log N)$ executions of the algorithm. This improves upon the $O(N)$ sequential restarts of~\cite{ChangEtAl2018,HaverlyLopez2021,SanyalBhaduriEtAl2022}, matching the binary-search strategy of~\cite{DiasZateska2021}, although it requires more measurements than the $O(1)$ dynamic-register approach of Han et al.~\cite{HanZhengChenWang2025}. Combining the dynamic threshold register of~\cite{HanZhengChenWang2025} with our $O(N)$-qubit QPE core is a natural direction for further improvement.

Finally, as pointed out in Section~\ref{subsec:complexity} and in~\cite{PelayoQPE2024}, relying on Quantum Phase Estimation introduces a hardware precision requirement that must be taken into account. Distinguishing $M$ edges or hyperedges exactly requires a base rotation angle $\theta = 2\pi / 2^{\lceil \log_2 M \rceil + 1} = \Theta(1/M)$. In the case of graphs, since $|E(\bar{G})| \le N(N-1)/2$, the smallest phase angle scales inverse-polynomially as $\Omega(1/N^2)$, which falls within standard fault-tolerant gate-synthesis regimes. In the case of hypergraphs with an exponential number of hyperedges ($M \in O(2^N)$), the angle $\theta$ scales as $\Omega(2^{-N})$ and hyperedges of size $\Theta(N)$ require $O(N)$-controlled phase gates. In that regime, our construction effectively shifts the exponential complexity from the number of qubits to the precision of the phase rotations and circuit depth.

\section{Conclusion}\label{sec:conclusion}

In this paper, we have shown how the GEO, originally introduced as a quantum phase estimation invariant for graph isomorphism~\cite{PelayoQPE2024}, can be used to build an ancilla-efficient Grover oracle for the Maximum Clique Problem, and generalized into a HEO to address the MHSP and the dualization of positive Boolean functions. By counting induced edges (or covered hyperedges) in the phase of the basis states and checking vertex subset cardinalities through a two's-complement weight register, both constructions avoid the need for explicit edge or hyperedge ancilla registers.

For graphs of $N$ vertices, the resulting MCP oracle requires $N + \lceil \log_2 |E(\bar{G})| \rceil + \lceil \log_2 N \rceil + 3 = O(N)$ qubits, reducing the $O(N^2)$ space complexity of existing gate-model Grover algorithms to linear while requiring $O(|E(\bar{G})|\log|E(\bar{G})| + N\log N)$ gates per oracle call. Our numerical experiments in Qiskit on graphs from $N=5$ to $N=10$ confirm that the compute--mark--uncompute sequence amplifies the target clique states as predicted by the theoretical bounds. For hypergraphs with $N$ vertices and $M$ hyperedges, the HEO uses inclusion--exclusion phase cancellation to count covered hyperedges in superposition, yielding an MHSP oracle that requires $O(N + \log M)$ qubits. This remains $O(N)$ even when the hypergraph has an exponential number of hyperedges, $M \in O(2^N)$. Moreover, measuring the phase register of $QPE_{\mathcal{H}}$ under a uniform superposition provides an approximation of the total number of hitting sets, which directly addresses the dualization problem for positive Boolean functions in PIDNF~\cite{Mezzini2024}.

Several lines of future work follow naturally from these results. First, it would be interesting to integrate the dynamic threshold register of Han et al.~\cite{HanZhengChenWang2025} with the GEO and HEO circuits so as to reduce the number of measurements for finding the optimum $k$ from $O(\log N)$ to $O(1)$ while keeping the $O(N)$ qubit bound. Second, the phase-encoding mechanism can be extended to vertex-weighted and hyperedge-weighted variants of both problems by encoding integer weights into the phase angles~\cite{Mills19}. Finally, combining the readout of the $QPE_{\mathcal{H}}$ phase register with Quantum Amplitude Estimation (QAE) could provide a quadratic speed-up in estimating the number of hitting sets for positive Boolean function dualization.

\section*{Acknowledgments}
This publication continues work supported by the I+D+i grant PID2021-122215NB-C32 (MICIU/AEI/10.13039/501100011033 and ERDF/EU), the UCLM group research grant 2022-GRIN-34113 (co-financed with ERDF Funds), the Region of Madrid project FORTE-CM (S2018/TCS-4314, co-funded by EIE/EU), and QSIMOV Quantum Computing S.L.\ (project 220426UCTR).

\bibliographystyle{plain}
\bibliography{References}

@article{Mezzini2024,
  author  = {Mezzini, Mauro and Cuartero Gomez, Fernando and Gonzalez, Jose Javier Paulet and de la Cruz Calvo, Hernan Indibil and Pascual, Vicente and L. Pelayo, Fernando},
  title   = {Polynomial quantum computing algorithms for solving the dualization problem for positive Boolean functions},
  journal = {Quantum Machine Intelligence},
  year    = {2024},
  volume  = {6},
  number  = {2},
  pages   = {71},
  month   = {oct},
  doi     = {10.1007/s42484-024-00202-y},
  url     = {https://doi.org/10.1007/s42484-024-00202-y},
  issn    = {2524-4914}
}

@Inbook{Karp1972,
author="Karp, Richard M.",
editor="Miller, Raymond E.
and Thatcher, James W.
and Bohlinger, Jean D.",
title="Reducibility among Combinatorial Problems",
bookTitle="Complexity of Computer Computations: Proceedings of a symposium on the Complexity of Computer Computations, held March 20--22, 1972, at the IBM Thomas J. Watson Research Center, Yorktown Heights, New York, and sponsored by the Office of Naval Research, Mathematics Program, IBM World Trade Corporation, and the IBM Research Mathematical Sciences Department",
year="1972",
publisher="Springer US",
address="Boston, MA",
pages="85--103",
isbn="978-1-4684-2001-2",
doi="10.1007/978-1-4684-2001-2_9",
url="https://doi.org/10.1007/978-1-4684-2001-2_9"
}

@article{Marino2026,
  author    = {Marino, Raffaele and Buffoni, Lorenzo and Zavalnij, Bogdan},
  title     = {A short review on the maximum clique problem algorithms with classical, AI, and quantum methods},
  journal   = {Communications Physics},
  year      = {2026},
  month     = {jul},
  volume    = {9},
  number    = {1},
  pages     = {239},
  issn      = {2399-3650},
  doi       = {10.1038/s42005-026-02606-7},
  url       = {https://doi.org/10.1038/s42005-026-02606-7}
}

@article{CaludeDinneenHua2017TCS,
  author  = {Calude, Cristian S. and Dinneen, Michael J. and Hua, Richard},
  title   = {QUBO Formulations for the Graph Isomorphism Problem and Related Problems},
  journal = {Theoretical Computer Science},
  volume  = {701},
  pages   = {54--69},
  year    = {2017},
  publisher = {Elsevier},
  doi     = {10.1016/j.tcs.2017.04.016},
  url     = {https://doi.org/10.1016/j.tcs.2017.04.016}
}

@article{HuaDinneen2020ImprovedQUBO,
  author    = {Hua, Richard and Dinneen, Michael J.},
  title     = {Improved QUBO Formulation of the Graph Isomorphism Problem},
  journal   = {SN Computer Science},
  volume    = {1},
  number    = {19},
  year      = {2020},
  publisher = {Springer Nature Singapore},
  doi       = {10.1007/s42979-019-0020-1},
  url       = {https://doi.org/10.1007/s42979-019-0020-1}
}

@article{Mills19,
  title = {Quantum invariants and the graph isomorphism problem},
  author = {Mills, P. W. and Rundle, R. P. and Samson, J. H. and Devitt, Simon J. and Tilma, Todd and Dwyer, V. M. and Everitt, Mark J.},
  journal = {Phys. Rev. A},
  volume = {100},
  issue = {5},
  pages = {052317},
  numpages = {11},
  year = {2019},
  publisher = {American Physical Society},
  doi = {10.1103/PhysRevA.100.052317},
  url = {https://link.aps.org/doi/10.1103/PhysRevA.100.052317}
}

@article{PelayoQPE2024,
  author  = {Pelayo, Fernando L. and Pascual Fidalgo, Vicente and de la Cruz Calvo, Hernan I. and Llana Diaz, Luis F. and Paulet Gonzalez, Jose J. and Mezzini, Mauro and Cuartero Gomez, Fernando},
  title   = {Towards Graph Isomorphism Characterisation Through Quantum Phase Estimation},
  journal = {ACM Transactions on Quantum Computing},
  year    = {2024},
  note    = {Under review; companion paper},
}

@inproceedings{Grover96,
  author    = {Grover, Lov K.},
  title     = {A Fast Quantum Mechanical Algorithm for Database Search},
  booktitle = {Proceedings of the Twenty-Eighth Annual ACM Symposium on Theory of Computing (STOC '96)},
  year      = {1996},
  pages     = {212--219},
  publisher = {ACM},
  doi       = {10.1145/237814.237866},
}

@article{ChatterjeeMohtashimKundu21,
author = {Chatterjee, Turbasu and Mohtahsim, Shah Ishmam and Kundu, Akash},
year = {2021},
month = {11},
pages = {},
title = {On The Variational Perspectives To The Graph Isomorphism Problem},
doi = {10.48550/arXiv.2111.09821},
journal = {}
}

@misc{ChildsFarhiGoldstoneGutmann2000,
  author = {Childs, Andrew M. and Farhi, Edward and Goldstone, Jeffrey and Gutmann, Sam},
  title  = {Finding cliques by quantum adiabatic evolution},
  year   = {2000},
  note   = {arXiv:quant-ph/0012104},
}

@inproceedings{PelofskeHahnDjidjev2019,
  author    = {Pelofske, Elijah and Hahn, Georg and Djidjev, Hristo N.},
  title     = {Solving large maximum clique problems on a quantum annealer},
  booktitle = {Quantum Technology and Optimization Problems (QTOP 2019)},
  series    = {Lecture Notes in Computer Science},
  volume    = {11413},
  pages     = {123--135},
  year      = {2019},
  publisher = {Springer},
}

@article{Bojic2012,
  author  = {Boji\'c, Ana},
  title   = {Quantum algorithm for finding a maximum clique in an undirected graph},
  journal = {Journal of Information and Organizational Sciences},
  volume  = {36},
  pages   = {91--98},
  year    = {2012},
}

@article{ChangEtAl2018,
  author  = {Chang, Weng-Long and Yu, Qi and Li, Zhaokai and Chen, Jian and Peng, Xinhua and Feng, Mang},
  title   = {Quantum speedup in solving the maximal-clique problem},
  journal = {Physical Review A},
  volume  = {97},
  pages   = {032344},
  year    = {2018},
  doi     = {10.1103/PhysRevA.97.032344},
}

@article{MetwalliLeGallVanMeter2020,
  author  = {Ayman Metwalli, Sara and Le Gall, Fran\c{c}ois and Van Meter, Rodney},
  title   = {Finding Small and Large $k$-Clique Instances on a Quantum Computer},
  journal = {IEEE Transactions on Quantum Engineering},
  volume  = {1},
  pages   = {1--11},
  year    = {2020},
  doi     = {10.1109/TQE.2020.3045692},
}

@article{SanyalBhaduriEtAl2022,
  author  = {Sanyal (Bhaduri), Arpita and Saha, Amit and Saha, Debasri and Saha, Banani and Chakrabarti, Amlan},
  title   = {Circuit design for clique problem and its implementation on quantum computer},
  journal = {IET Quantum Communication},
  volume  = {3},
  number  = {1},
  pages   = {30--49},
  year    = {2022},
  doi     = {10.1049/qtc2.12029},
  note    = {Preprint: arXiv:2004.10596},
}

@misc{SanyalBhaduriQudits2023,
  author = {Sanyal (Bhaduri), Arpita and Saha, Amit and Saha, Banani and Chakrabarti, Amlan},
  title  = {Robust Quantum Circuit for Clique Problem with Intermediate Qudits},
  year   = {2022},
  note   = {arXiv:2211.07947; published in Procedia Computer Science (2023)},
}

@inproceedings{DiasZateska2021,
  author    = {Dias, Matheus Giovanni and Zateska, Leyza Baldo},
  title     = {Grover's quantum algorithm and the maximum clique problem},
  booktitle = {Proceedings of the 2021 Workshop de Pesquisa em Computa\c{c}\~ao dos Campos Gerais (WPCCG)},
  pages     = {45--52},
  year      = {2021},
  address   = {Ponta Grossa, Brazil},
  publisher = {UTFPR},
}

@inproceedings{HaverlyLopez2021,
  author    = {Haverly, Andrew and L\'opez, Sonia},
  title     = {Implementation of Grover's Algorithm to Solve the Maximum Clique Problem},
  booktitle = {2021 IEEE Computer Society Annual Symposium on VLSI (ISVLSI)},
  pages     = {441--446},
  year      = {2021},
  publisher = {IEEE},
}

@mastersthesis{Haverly2021Thesis,
  author = {Haverly, Andrew R.},
  title  = {A Comparison of Quantum Algorithms for the Maximum Clique Problem},
  school = {Rochester Institute of Technology},
  year   = {2021},
  note   = {Thesis, comparing qubit gate-model and boson-sampling implementations},
}

@misc{HanZhengChenWang2025,
  author = {Han, Wenmin and Zheng, Shiqi and Chen, Peian and Wang, Yukun},
  title  = {Efficient Maximum Clique Detection via Grover's Algorithm with Real-time Global Size Tracking},
  year   = {2025},
  note   = {arXiv:2509.01261},
}

@inproceedings{McCreesh2019,
  author    = {McCreesh, Ciaran and Pettersson, William and Prosser, Patrick},
  title     = {Understanding the empirical hardness of random optimisation problems},
  booktitle = {Principles and Practice of Constraint Programming (CP 2019)},
  pages     = {333--349},
  publisher = {Springer},
  year      = {2019}
}

@article{SanSegundo2023,
  author  = {San Segundo, Pablo and Furini, Fabio and {\'A}lvarez, David and Pardalos, Panos M.},
  title   = {{CliSAT}: A new exact algorithm for hard maximum clique problems},
  journal = {European Journal of Operational Research},
  volume  = {307},
  pages   = {1008--1025},
  year    = {2023}
}

@article{DurrHoyer1996,
  author  = {D{\"u}rr, Christoph and H{\o}yer, Peter},
  title   = {A quantum algorithm for finding the minimum},
  journal = {arXiv preprint quant-ph/9607014},
  year    = {1996}
}

@article{LiCTQW2021,
  author  = {Li, Xiaoyu and Wu, Meng and Chen, Hanwu and Liu, Zhihao},
  title   = {Algorithm for finding the maximum clique based on continuous time quantum walk},
  journal = {Quantum Information and Computation},
  volume  = {21},
  pages   = {59--79},
  year    = {2021}
}

@article{LiSeidel2024,
  author  = {Li, Xiaoyu and Chen, X. and Hu, S. and Xu, J. and Liu, Zhihao},
  title   = {A parameter-independent algorithm of finding maximum clique with {Seidel} continuous-time quantum walks},
  journal = {iScience},
  volume  = {27},
  pages   = {108953},
  year    = {2024}
}

@article{YinPhotonic2023,
  author  = {Yin, Xiao-Feng and others},
  title   = {Solving independent set problems with photonic quantum circuits},
  journal = {Proceedings of the National Academy of Sciences (PNAS)},
  volume  = {120},
  pages   = {e2212323120},
  year    = {2023}
}

@article{Krpan2025,
  author  = {Krpan, A. and Povh, Janez and Pucher, D.},
  title   = {Quantum computing and the stable set problem},
  journal = {Optimization Methods and Software},
  volume  = {40},
  pages   = {837--870},
  year    = {2025}
}

@article{okamoto2020finding,
  title={Finding a maximum common subgraph from molecular structural formulas through the maximum clique approach combined with the Ising model},
  author={Okamoto, Yasuharu},
  journal={ACS omega},
  volume={5},
  number={22},
  pages={13064},
  year={2020}
}

@article{rozman2024enhanced,
  title={Enhanced molecular docking: Novel algorithm for identifying highest weight k-cliques in weighted general and protein-ligand graphs},
  author={Rozman, Kati and Ghysels, An and Zavalnij, Bogdan and Kunej, Tanja and Bren, Urban and Jane{\v{z}}i{\v{c}}, Du{\v{s}}anka and Konc, Janez},
  journal={Journal of molecular structure},
  volume={1304},
  pages={137639},
  year={2024},
  publisher={Elsevier}
}

@article{boginski2005statistical,
  title={Statistical analysis of financial networks},
  author={Boginski, Vladimir and Butenko, Sergiy and Pardalos, Panos M},
  journal={Computational statistics \& data analysis},
  volume={48},
  number={2},
  pages={431--443},
  year={2005},
  publisher={Elsevier}
}

@article{douik2020tutorial,
  title={A tutorial on clique problems in communications and signal processing},
  author={Douik, Ahmed and Dahrouj, Hayssam and Al-Naffouri, Tareq Y and Alouini, Mohamed-Slim},
  journal={Proceedings of the IEEE},
  volume={108},
  number={4},
  pages={583--608},
  year={2020},
  publisher={IEEE}
}

@article{lu2023atomic,
  title={On atomic cliques in temporal graphs: Y. Lu et al.},
  author={Lu, Yajun and Miao, Zhuqi and Sahraeian, Parisa and Balasundaram, Balabhaskar},
  journal={Optimization Letters},
  volume={17},
  number={4},
  pages={813--828},
  year={2023},
  publisher={Springer}
}

\end{document}